\documentclass[aps,prb,amsmath,amssymb,twocolumn,10pt,footinbib,showpacs]{revtex4-1}
\usepackage{graphicx}
\usepackage{graphics,color}
\usepackage{hyperref}
\usepackage{hypernat}

\newcommand{\bs}{\;\;\;\;\;}

\newcommand{\ve}{\mathbf}

\newcommand{\D}{{\rm d}}
\newcommand{\1}{t}
\newcommand{\2}{z}

\usepackage{color}

\begin{document}

\title{Carbon nanotubes in electric and magnetic fields}
\author{Jelena Klinovaja}
\author{Manuel J. Schmidt}
\author{Bernd Braunecker}
\author{Daniel Loss}
\affiliation{Department of Physics, University of Basel, Klingelbergstrasse 82, 4056 Basel,
Switzerland}
\date{\today}


\begin{abstract}
We derive an effective low-energy theory for metallic (armchair and non-armchair) single-wall nanotubes in the presence of an electric field perpendicular to the nanotube axis, and in the presence of magnetic fields, taking into account spin-orbit interactions and screening effects on the basis of a microscopic tight binding model. 
The interplay between electric field and spin-orbit interaction allows us to tune armchair nanotubes into a helical conductor in both Dirac valleys. Metallic non-armchair nanotubes are gapped by the surface curvature, yet helical conduction modes can be restored in one of the valleys by a magnetic field along the nanotube axis. Furthermore, we discuss electric dipole spin resonance in carbon nanotubes, and find that the Rabi frequency shows a pronounced dependence on the momentum along the nanotube.
\end{abstract}

\maketitle

\section{Introduction}

The last two decades have seen remarkable progress in the experimental techniques to fabricate and analyze high purity carbon nanotubes (CNTs).\cite{tans_cnt_exp_1997, cobden_cnt_exp_2002, liang_cnt_exp_2002, minot_cnt_exp_soi_2004, jarillo-herrero_cnt_exp_nature_2004, jarillo-herrero_cnt_exp_2005, graeber_cnt_exp_2006} This progress has paved the way for using CNTs for electron and, in particular, electron spin based effects that are of interest for quantum information processing and spintronics.\cite{bulaev_soi_cnt_2008, kuemmeth_exp_cnt_2008,  churchill_exp_cnt_2009, steele_exp_cnt_2009, aurich_exp_cnt_2010,jespersen_soi_cnt_exp_2011, holm_cnt_2008, jarillo-herrero_cnt_exp_nature_2005, sapmaz_cnt_exp_2005, gunnarsson_cnt_exp_2008, sahool_cnt_exp_2008, kuemmeth_review_cnt_2010, palyi_spin_valley_blockade_2010,rudner_cnt_qd_spin_relax_2010,flensberg_bend_cnt_spin_control_2010,weiss_soi_2010} For the latter, the spin-orbit interaction (SOI) plays a significant role as it allows the spin manipulation by electric fields. The purity of CNTs has by now advanced so far that indeed SOI effects can be observed.\cite{kuemmeth_exp_cnt_2008, churchill_exp_cnt_2009, steele_exp_cnt_2009, aurich_exp_cnt_2010,jespersen_soi_cnt_exp_2011}

In this paper, we investigate the SOI in metallic single-wall CNTs in the presence of external electric and magnetic fields. We provide an extensive discussion of electric field screening in CNTs and of resonant spin transitions via ac electric fields, namely, electric dipole spin resonance (EDSR). Our starting point is a tight binding description including all second shell orbitals of the carbon atoms, from which we derive an effective low-energy band theory. 

In the metallic regime, SOI effects by an external electric field can only be expected if the field is applied perpendicularly to the CNT axis. Without the electric field, the SOI in a CNT has been studied before\cite{ando_soi_cnt_2000, huertas-hernando_soi_cnt_2006, chico_soi_numerics_2009, izumida_soi_cnt_2009, jeong_soi_cnt_2009, white_soi_cnt_2009} and is the result of the orbital mixing caused by the curvature of the graphene sheet wrapped up into a cylinder. This leads to different geometric conditions than for flat graphene, which has accordingly a different SOI.\cite{kane_soi_graphene_2005, min_soi_2006} It is also very different from the SOI found in typical one-dimensional semiconductor wires because of the rotational symmetry of the CNT. In a semiconductor wire, the usual Rashba-SOI with a well-defined spin-precession axis arises from the specific asymmetric electric environment caused by the confining potentials. Such an axis is absent in the rotationally invariant CNT unless it is reintroduced by the application of an external transverse electric field. As we discuss in detail below, this leads indeed to a Rashba-like SOI. However, in contrast to the semiconductors, the CNTs have a hexagonal lattice structure with two carbon atoms per unit cell. The resulting band structure has a nearly vanishing density of states at the charge neutrality point (Dirac point), and roughly a linearly increasing density of states at low energies away from this point. This means, first, that the high energy states affect the low energy properties in a more pronounced way than in a semiconductor. Second, the screening of the external electric field becomes less effective, and the distribution of charges on the tube surface can become complicated. As the screened field affects the low-energy physics as well, it requires specific investigation. It turns out here that for a quantitative understanding we need to start indeed from the full band structure based on the lattice description, and not from the effective low-energy Dirac theory.

The combination of these effects allows us to understand how precisely SOI and especially the SOI parts induced by the external electric field affect the system properties. The larger number of external and internal degrees of freedom as compared to semiconductor wires, such as field strength, sublattice or Dirac valley index, CNT radius, chirality, and chemical potential, allows for an extensive tunability of the SOI-induced system properties. For instance, in armchair CNTs it allows us to obtain in both Dirac valleys helical, spin-filtered conduction modes, in which opposite spins are transported in opposite directions. Similar helical states occur in SOI-split quantum wires\cite{streda_2003,pershin_2004,devillard_2005,zhang_2006,sanchez_2008, birkholz_2009,quay_2010,braunecker_wire_2010} and at the edges of topological insulators,\cite{hasan_topological_2010} and can be used, for instance, as spin filters\cite{streda_wire_2010,braunecker_wire_2010} or as Cooper-pair splitters.\cite{sato_cooper_pair_2010} In addition, if such a helical conductor is brought into proximity to a superconductor, it allows for the realization of Majorana end states. This has attracted much attention very recently,\cite{lutchyn_majorana_wire_2010, oreg_majorana_wire_2010,potter_majoranas_2011,alicea_majoranas_2010, suhas_arxiv_2011} mainly because these Majorana states may be used as fundamental quantum states for topological quantum computation.\cite{nayak_top_qc_majorana_2008}

It should be noted that in semiconductor quantum wires the helical modes are realized only under conditions of an external magnetic field, with the consequence that time-reversal symmetry is broken and the spins at opposite conduction band branches are not truly antiparallel. In contrast, in armchair CNTs the helical modes can be obtained in an all-electric setup, and they are perfectly polarized in the sense that only the spin expectation value $S^y$ in the direction perpendicular to the electric field and the CNT axis is nonzero, even though we generally have $|\langle S^y\rangle| < 1$ within the helical bands. 

For metallic but non-armchair CNTs, two more terms appear in the low-energy Hamiltonian. First, an orbital curvature induced term, which opens gaps at the remaining helical zero-energy modes. Second, a SOI term, which plays the role of an effective Zeeman field transverse to the $S^y$ polarization, yet with opposite sign in each Dirac valley. While these two terms primarily destroy the helical conduction modes, their presence can be turned to an advantage: by applying a magnetic field along the CNT axis, the effective Zeeman field in one of the valleys can be suppressed, while it doubles the gap in the other valley. Hence the helical conduction modes are restored to high quality in one of the valleys, while conduction in the other valley is suppressed completely.

To conclude, we provide a microscopic description of electric dipole spin resonance\cite{kato_edsr_2003,golovach_edsr_2006,nowack_edsr_2007,bulaev_edsr_2007, bulaev_soi_cnt_2008, laird_edsr_2009} (EDSR) in CNTs. An ac electric field perpendicular to the CNT axis couples to the electronic spin via the SOI. However, this coupling comprises an additional sublattice coupling that is absent in semiconductor setups. It causes a significant momentum dependence of the resonant Rabi frequency of the EDSR experiment. The further application of a static electric field, perpendicular to the CNT axis and to the ac field, lifts the spin degeneracy of the bands in an way analogous to a static Zeeman field. This allows us to propose an all electric realization of Rabi resonance experiments, in which the electric fields replace the static and time-dependent magnetic fields.

The paper is structured as follows. In Sec. \ref{section_model} we introduce the tight-binding model of the hexagonal carbon lattice including SOI and the effect of electric and magnetic fields. The $\pi$ and $\sigma$ bands are discussed in Sec. \ref{pi_sigma_section}, and we identify the important $\pi\sigma$ band hybridization couplings that have an impact on the low-energy physics. This allows us to derive the effective low-energy theory for the $\pi$ bands in Sec. \ref{sect_effective_hamiltonian}. The partial screening of the external electric field is investigated in Sec. \ref{section_screening}, where we discuss in particular the validity of linear response and the influence of the $\sigma$ bands. The implications of the effective low-energy theory are analyzed in Sec. \ref{sect_analysis_eff_theory}, with special focus on helical modes, valley suppression and the role of further external magnetic fields. Turning to the dynamical response of the CNTs, in Sec. \ref{section_edsr}, we investigate resonant spin transitions by an ac electric field. The final section \ref{section_conclusions} contains our conclusions.

\section{Model\label{section_model}}
\begin{figure}[thp!]
\centering
\includegraphics[width=180pt]{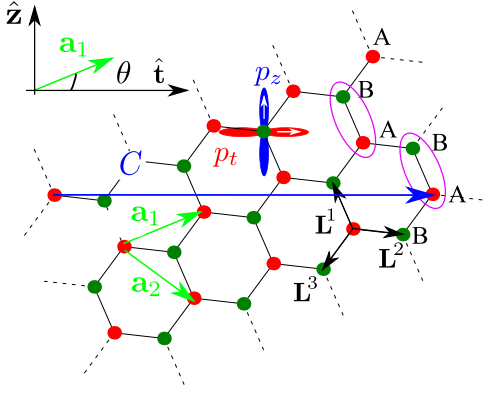}
\caption{(Color online) Definition of the honeycomb lattice vectors and the direction of the $p_z$ and $p_t$ orbitals. The white arrows in the $p$ orbitals point into the direction where the wave function is positive. The ellipses indicate the unit cells consisting of A and B sublattice atoms. The $p$ orbital alignment is equal on both sublattices. Together with the $p_r$ orbital, which points out of the plane (not shown here), the directions of $p_r,p_t,p_z$ form a right-handed set of vectors.}
\label{fig_lattice_definitions}
\end{figure}

\begin{figure}[!ht]
\centering
\includegraphics[width=120pt]{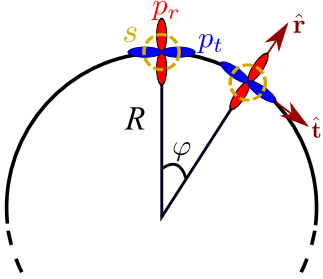}
\caption{(Color online) Cross section of a CNT. The $\hat{\ve z}$-direction is along the nanotube axis. The orientation of the orbitals $p_r,p_t$ as well as the local coordinate system $\hat {\ve r},\hat{\ve t}$ depends on the azimuthal angle $\varphi$. The $s$ orbital is indicated by the dashed circles.}
\label{fig_definitions_orbitals_on_the_surface}
\end{figure}

Our calculations are based on a tight-binding model describing the second shell orbitals of the carbon atoms in CNTs. \cite{dresselhaus_book} It takes into account the spin-orbit interaction and external electric and magnetic fields. The Hamiltonian we start from consists of four terms
\begin{equation}
H = H_{\rm hop} + H_{\rm SO} + H_{E} + H_{B}, \label{full_hamiltonian}
\end{equation}
where $H_{\rm hop}$ describes the hopping between neighboring tight-binding orbitals, $H_{\rm SO}$ describes the on-site spin-orbit interaction, and $H_{E}$ and $H_{B}$ describe external electric and magnetic fields, respectively. The Hamiltonian is expressed in second quantization by the electron annihilation operators $c_{\ve n, \zeta, \mu, \lambda}$, where the two-dimensional integer vector $\ve n=(n_1,n_2)$ labels the unit cells of a honeycomb lattice, $\zeta=\pm1$ labels the sublattice A/B, and $\lambda=\pm1$ labels the spin.  $\mu=s,p_r,p_t,p_z$ denotes the second shell orbitals with $\mu=s$ the $s$ orbital and $\mu=p_r,p_t,p_z$ the $p$ orbitals. The $p_r$ orbital is perpendicular to the surface of the CNT and essentially constitutes the $\pi$ band. The $p_{t,z}$ orbitals are tangential (see Figs. \ref{fig_lattice_definitions} and \ref{fig_definitions_orbitals_on_the_surface}) and, together with the $s$ orbital, build the $\sigma$ bands. These orbitals span a local coordinate system.

If the cylinder surface on which the carbon atoms are placed in a CNT is projected onto a plane, the graphene lattice with lattice vectors $\ve a_1 = a (\hat{\mathbf{t}} \cos\theta + \hat{\mathbf{z}} \sin\theta)$ and $\ve a_2 = a (\hat{\mathbf{t}}\cos(\pi/3-\theta) - \hat{\mathbf{z}} \sin(\pi/3-\theta))$ is recovered, where $\theta$ is the chiral angle of the nanotube, defined as the angle between the first unit vector $\ve a_1$ and the chiral vector $C$ (see Fig. \ref{fig_lattice_definitions}). In the following, we will also use the collective site index $i=(n_1,n_2,\zeta)$ to label the atom positions. The position of atom $i$ is then $\ve R_i=n_1\ve a_1+n_2\ve a_2+(1-\zeta)\ve L^1/2$. The three vectors
\begin{align}
\ve L^1 &= \frac a{\sqrt3} (-\hat{\mathbf{t}}\sin\theta + \hat {\mathbf{z}} \cos\theta), \\
\ve L^2 &= \frac a{\sqrt3} (\hat{\mathbf{t}}\cos(\pi/6-\theta) - \hat{\mathbf{z}}\sin(\pi/6-\theta)), \\
\ve L^3 &= \frac a{\sqrt3} (-\hat{\mathbf{t}}\cos(\pi/6+\theta) - \hat{\mathbf{z}}\sin(\pi/6+\theta)),
\end{align}
connect an A site with its three nearest neighbors on the B sublattice. These three B sites will be denoted by $B_{1,2,3}$ in the following. Note that we have chosen to define the model directly in a coordinate system which will be convenient for the subsequent calculations.

In the following, we describe each term in Eq. (\ref{full_hamiltonian}) and its derivation in detail.

\subsection{Hopping Hamiltonian}

\begin{figure}[thp!]
\centering
\includegraphics[width=\columnwidth]{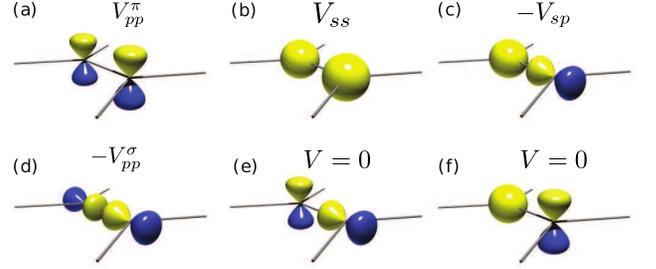}
\caption{(Color online) Hopping between atomic orbitals of neighboring carbon sites in the graphene limit. (a) - (d) show the orbital combinations with non-zero hopping amplitude. (e), (f) show hoppings which are forbidden by symmetry.}
\label{fig:bonds}
\end{figure}

\begin{figure}[h]
\begin{minipage}[h]{0.49\linewidth}
\center{\includegraphics[width=4cm]{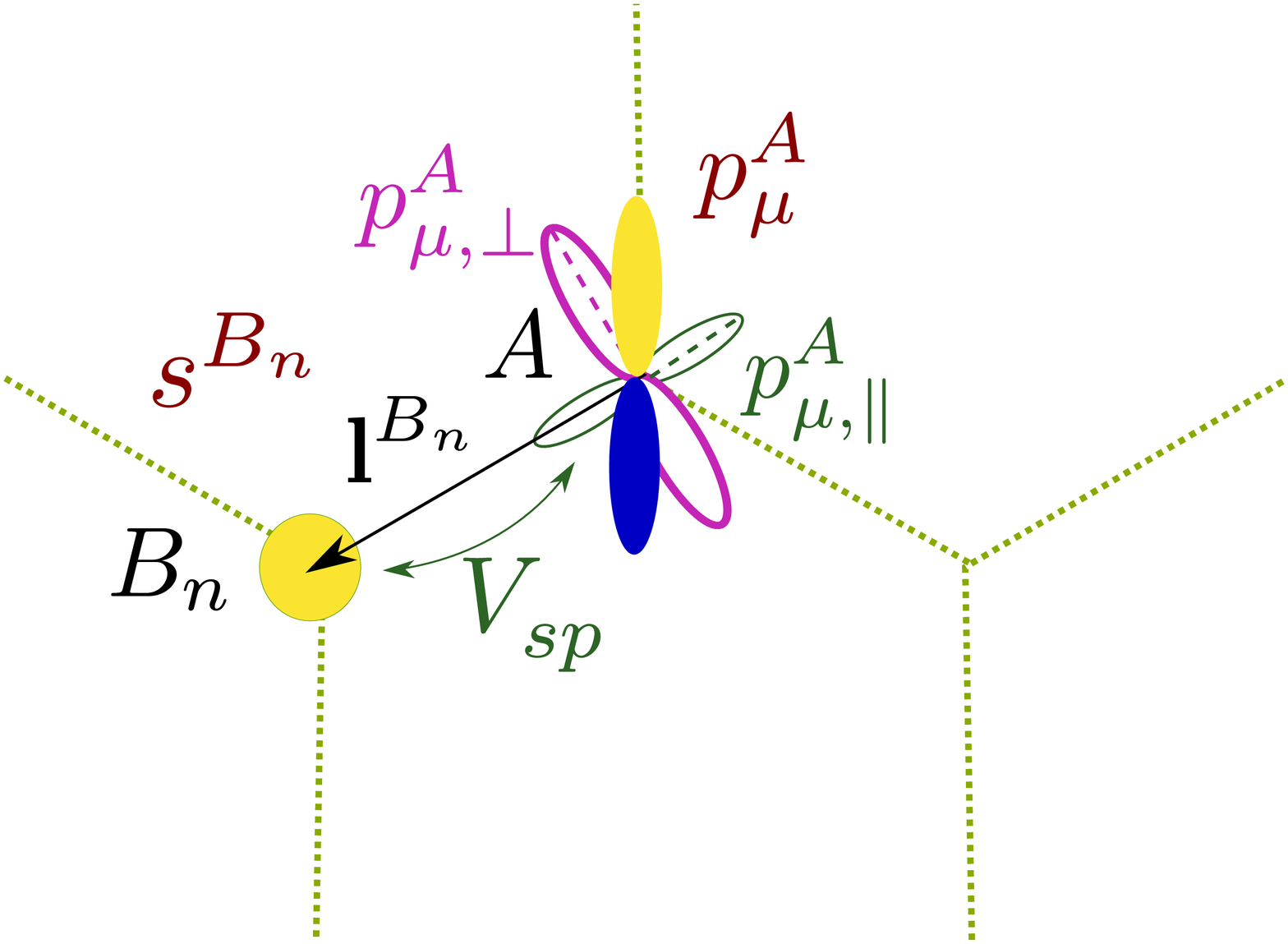}\\ a)}
\end{minipage}
\hfill
\begin{minipage}[h]{0.49\linewidth}
\center{\includegraphics[width=4cm]{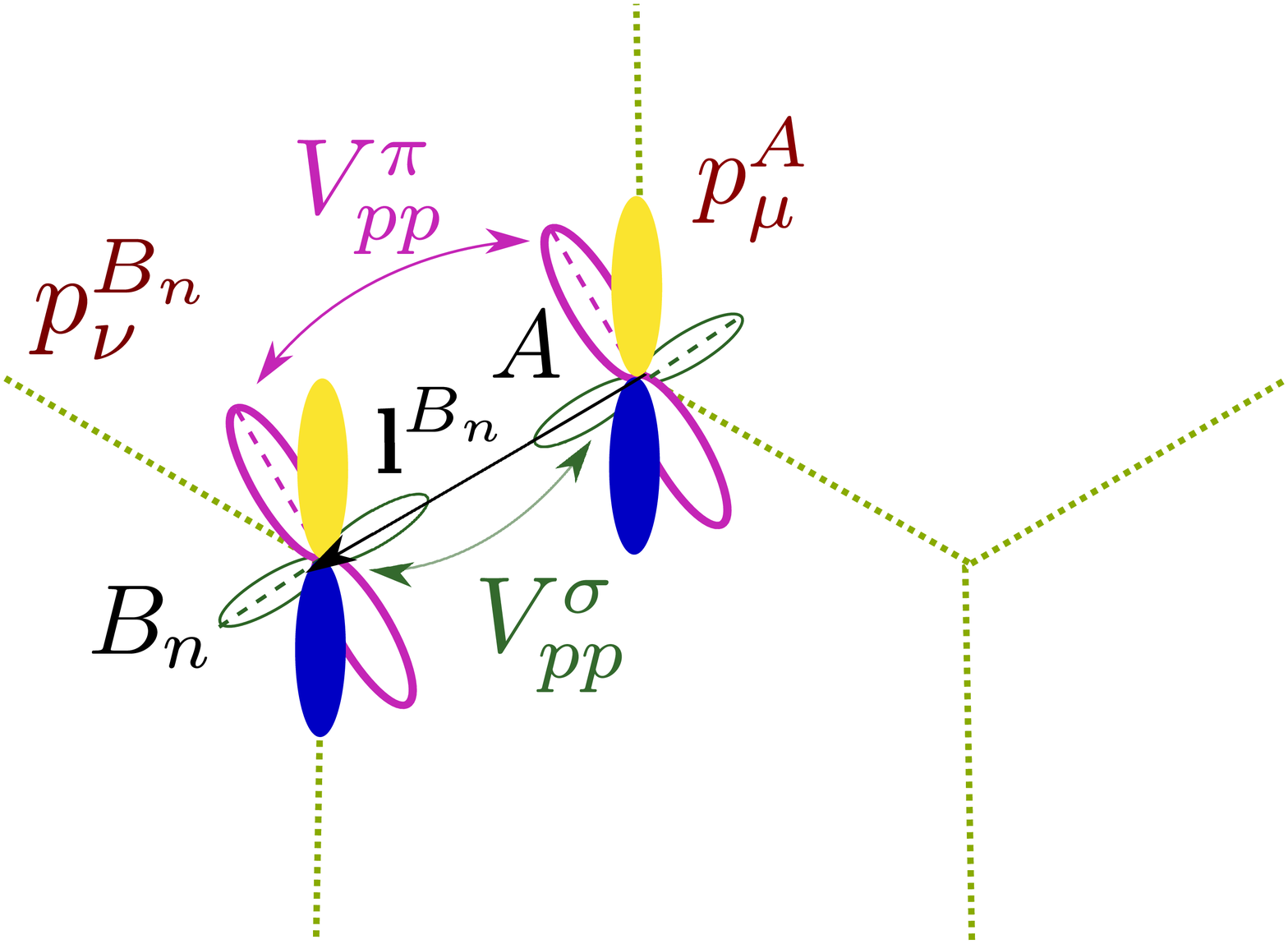}\\ b)}
\end{minipage}
\caption{(Color online) The hopping element between a) $p_\mu^A$ and $s^{B_n}$ b) $p_\mu^A$ and $p_\nu^{B_n}$ can be calculated by the decomposition of the $p_\mu^A$($p_\nu^{B_n}$) orbital into two parts: perpendicular to the bond $p^A_{\mu,\perp}$($p_{\nu,\perp}^{B_n}$) and parallel to the bond $p^A_{\mu,\parallel}$($p_{\nu,\parallel}^{B_n}$).}
\label{fig:hopping_explanations}
\end{figure}

The term $H_{\rm hop}$ describes the hopping between orbitals on neighboring carbon sites as well as the on-site orbital energies. We only take into account nearest neighbor hopping and assume the orbitals on neighboring carbon atoms to be orthogonal. The hopping Hamiltonian has the form
\begin{equation}
H_{\rm hop} = \sum_{\substack{\left<i,j\right>\\\mu,\mu',\lambda}} t_{\mu\mu'}^{ij} c^\dagger_{i\mu\lambda} c_{j\mu'\lambda} + \varepsilon_s \sum_{i,\lambda} c^\dagger_{is\lambda} c_{is\lambda},
\end{equation}
where $\left<i,j\right>$ runs over nearest neighbor sites, and $\varepsilon_s=-8.9$ eV is the orbital energy of the carbon $s$ orbitals relative to the $p$ orbital energy, The latter is set to zero in this paper. The amplitude $t_{\mu\mu'}^{ij}$ of an electron hopping from the orbital $\mu'$ on the $j$th site to the orbital $\mu$ on the $i$th site is a linear combination of the four fundamental hopping amplitudes \cite{dresselhaus_book} $V_{ss}=-6.8$ eV, $V_{sp}=5.6$ eV, $V_{pp}^\pi=-3.0$ eV, $V_{pp}^\sigma=5.0$ eV (see Fig. \ref{fig:bonds}) with coefficients depending on the relative orientation of the orbitals $\mu$ and $\mu'$.\cite{soi_schmidt_loss_2010} For the explicit calculation of $t_{\mu\mu'}^{ij}$, the $p$ orbitals are decomposed into components parallel and perpendicular to the $i-j$ bond. These components can be easily expressed in terms of scalar products (see Fig. \ref{fig:hopping_explanations}). If the direction of a $p$ orbital on the A atom is ${\bf p}_\mu^{A}$ and the direction of a $p$ orbital on the $B_n$ atom is ${\bf{p}}_\nu^{B_n}$, then the projections on the unit vector ${\bf{l}}^{n}={\bf{L}}^{n}/{\mid{\bf{L}}^{n}}\mid$ in the direction from atom $A$ to atom $B_n$  are
\begin{eqnarray}
\sigma_{\perp} &=& ({\bf p}_\mu^{A}\cdot {\bf{l}}^{n})({\bf{p}}_\nu^{B_n}\cdot {\bf{l}}^{n}),\\
\sigma_{\parallel}&=&{\bf p}_\mu^{A}\cdot{\bf{p}}_\nu^{B_n}-({\bf p}_\mu^{A}\cdot {\bf{l}}^{n})({\bf{p}}_\nu^{B_n}\cdot {\bf{l}}^{n}),
\end{eqnarray}
where indices $\{\mu, \nu\} = \{r,t,z\}$ describe $p$ orbitals and $n=\{ 1,2,3\}$ denotes the neighboring B atoms.
Therefore, the hopping matrix elements are
\begin{align}
t_{\mu \nu}^{AB_n} =& V_{pp}^\pi\left({\bf p}_\mu^{A}\cdot{\bf{p}}_\nu^{B_n} -({\bf{p}}_\mu^{A}\cdot {\bf{l}}^{n})({\bf{p}}_\nu^{B_n}\cdot {\bf{l}}^{n})\right)+\nonumber \\&V_{pp}^\sigma ({\bf{p}}_\mu^{A}\cdot {\bf{l}}^{n})({\bf{p}}_\nu^{B_n}\cdot {\bf{l}}^{n}),\label{eq:hab1}\\
t_{\mu s}^{AB_n}=& -V_{s p} ({\bf{p}}_\mu^{A}\cdot{\bf{l}}^{n}),\label{eq:hab2} \\
t_{s \nu}^{AB_n}=&  V_{s p}({\bf{p}}_\nu^{B_n}\cdot {\bf{l}}^{n}) \label{eq:hab3}.
\end{align}

\subsection{Spin-orbit interaction}

The spin-orbit interaction is modeled by projecting the Thomas Hamiltonian for a spherically symmetric atomic potential $V(r)$\cite{sakurai_book}
\begin{equation}
H_\mathrm{SO}^\mathrm{intr}=\frac{\hbar}{4m_e^2c^2}\frac1r \frac{\D V(r)}{\D r} \ve L\cdot \ve S
\label{eq:thomas}
\end{equation}
onto the subspace spanned by the second shell orbitals of each individual carbon atom. $\ve L$ is the electron angular momentum operator and $\ve S$ is the vector of the spin Pauli matrices with eigenvalues $\pm1$.

By symmetry, this projection can be reduced to a form depending on a single parameter only, defining the strength of the intrinsic SOI
\begin{eqnarray}
\Delta_{\rm SO}&=&\frac{\hbar^2}{4m_e^2c^2}\langle p_r^A\vert \left( \partial_t[\partial_r V]- \partial_r[\partial_t V]\right) \vert p_t^A\rangle.
\label{delta_so}
\end{eqnarray}
In the literature one finds a wide spread for the spin-orbit coupling constant $\Delta_{\rm SO}$ (see Refs. \onlinecite{serrano_soi_strength_abinitio_2000,min_soi_2006,guinea_soi_bilayer_2010}). Note also that there are different conventions for the spin operators used in the spin-orbit Hamiltonian. In some works, Pauli matrices with eigenvalues $\pm1$ are used while in other papers, the spin operators have eigenvalues $\pm\frac12$. In this paper, we always use Pauli matrices with eigenvalues $\pm1$. For the spin-orbit coupling constant one finds values between 3 meV\cite{min_soi_2006} and 20 meV.\cite{guinea_soi_bilayer_2010} In this work we follow Ref. \onlinecite{serrano_soi_strength_abinitio_2000} and use $\Delta_{\rm SO}= 6$ meV. The projected spin-orbit Hamiltonian reads
\begin{equation}
H_{\rm SO} =  i\Delta_{\rm SO} \sum_{\substack{i,\lambda,\lambda'\\\mu,\nu,\eta}} \varepsilon^{\mu\nu\eta} c^\dagger_{i\mu\lambda} S^\nu_{\lambda\lambda'} c_{i\eta\lambda'}.\label{hamiltonian_spin_orbit}
\end{equation}
Here, $\mu,\eta=p_r,p_t,p_z$ run over the $p$ orbitals only. $\varepsilon^{\mu\nu\eta}$ is the Levi-Civita symbol. The index $\nu=r,t,z$ labels the spin components in the local coordinate system, i.e., $S^r = S^x \cos\varphi_i  + S^y\sin\varphi_i$, $S^t = S^y \cos\varphi_i - S^x \sin\varphi_i$, with $\varphi_i$ the azimuthal angle of site $i$ (see Fig. \ref{fig_definitions_orbitals_on_the_surface}) and $S^{x,y,z}$ the spin Pauli matrices (with eigenvalues $\pm 1$). The spin-orbit energies emerging from $H_{\rm SO}$ and from the curvature effects are found to be much larger than the spin-orbit energies due to the $d$-orbitals, \cite{gmitra_spin_orbit_d_orbitals_2009} allowing us to neglect the latter.

\subsection{Electric fields\label{sec_definitions_el_field}}

In this section we introduce the Hamiltonian describing a homogenous external electric field $\ve E$ perpendicular to the CNT axis. Due to screening effects, local fields with complicated spatial dependence are generated by the rearrangement of the electron density on the lattice and have to be taken into account in principle. However, it turns out that none of these nontrivial contributions affect the physics in an essentialy way. 

We start from the most general electrostatic potential $\phi_{\rm tot}(\ve r)=\phi_{\rm ext}(\ve r)+\phi_{\rm ind}(\ve r)$, which is the sum of the potential $\phi_{\rm ext}$ coming from the homogeneous external field and the generally complicated potential $\phi_{\rm ind}$ from screening effects. The matrix elements $\left<\mu,i\right|\phi(\ve r)\left|\mu',j\right>$ between the orbitals of two different carbon atoms $i\neq j$ are much smaller than the typical hopping elements ($\sim $ eV) so that we can safely neglect them and consider only on-site effects of the electric potential.

The averaged potential $\phi_{\mu \mu}(\ve R_i) =\left<\mu,i\right| \phi_{\rm tot}(\ve r) \left|\mu,i\right> $ for the $\mu$ orbital on site $i$ changes the electrostatic energy of an electron in this orbital. Note that $\ve R_i$ is the position of the $i$th carbon atom in three-dimensional space. In the following, we assume that the dependence of $\phi_{\mu\mu}(\ve R_i)$ on the orbital $\mu$ is negligible, i.e., $\phi_{\rm tot}(\ve R_i) \simeq \phi_{\mu\mu}(\ve R_i)$. Furthermore,
we show in Sec. \ref{section_screening} that even with the induced potential $\phi_{\rm ind}$ included, $\phi_{\rm tot}$ essentially depends on the azimuthal angle $\varphi_i$ similar to the case of a homogeneous field. Therefore, the diagonal matrix elements of the electrostatic potential are $\phi(\ve R_i)\simeq\phi_{\rm tot}(\varphi_i)\propto\cos\varphi_i$ and give rise to the Hamiltonian
\begin{equation}
H_E^{(1)} = \sum_{i,\mu,\lambda} \phi_{\rm tot}(\varphi_i) c^\dagger_{i\mu\lambda} c_{i\mu\lambda}.
\end{equation}
For sufficiently small CNT radii $R$ and electric fields the total potential on the surface of the CNT is well approximated by $\phi_{\rm tot} (\varphi_i) \simeq  e E^* R \cos\varphi_i$, where $E^*<   |\ve E|$ is the screened electric field and $e$ is the electron charge.

We note that, because $\phi$ varies on the scale of the spatial extent of the orbital wave function and breaks the lattice symmetries, in general, the matrix elements between the different orbitals $\phi_{\mu \mu'}(\ve R_i) =\left<\mu,i\right| \phi(\ve r) \left| \mu',i\right> \neq 0$ and transitions between orthogonal orbitals $\mu,\mu'$ on the same carbon atom are generated. We call this the $\mu$-$\mu'$ transitions in the following. The potential of a homogeneous external field alone gives rise to $s$-$p_r$ and $s$-$p_t$ transitions. But the complicated additional induced potential $\phi_{\rm ind}(\ve r)$ gives also rise to coupling between other orbitals. For this work, the $s$-$p_r$ transition is most important because of two reasons: 1) It is the only transition directly coupling $\pi$ and $\sigma$ bands, thus giving rise to a first order effect in the $s$-$p_r$ coupling strength. 2) Its strength is determined by the unscreened field $E$ and not by the screened $E^*<E$. Indeed, the induced potential $\phi_{\rm ind}$ drops out, i.e., $\left<p_r\right|\phi_{\rm ind}(\ve r)\left|s\right>\simeq0$, as $\phi_{\rm ind}$ is approximately an even function in the radial coordinate $r$ about the tube radius $r=R$. The Hamiltonian describing the $s$-$p_r$ transition is
\begin{equation}
H_{E}^{(2)} = - e E \xi_0 \cos(\varphi_i) c^\dagger_{ip_r\lambda} c_{is\lambda} + {\rm H.c.},
\end{equation}
with the strength of the transitions being characterized by the integral
\begin{equation}
\xi_0 = - \int \D^3 \ve r \, \psi_{2s}^*(\ve r) \,z\, \psi_{2p_z}(\ve r)=\frac{3 a_B}{Z}, \label{spr_integral}
\end{equation}
where $\psi_{2s}$ and $\psi_{2p_z}$ are the hydrogenic wave functions of the second shell atomic orbitals, $a_B$ is the Bohr radius, and $Z$ is the effective nuclear charge, which for the second shell in carbon is $Z\simeq 3.2$. From this we obtain $\xi_0 \simeq 0.5\,\text\AA$. Note that our value for $\xi_0$ is a rather conservative estimate. It is roughly four times smaller than what has been assumed in Ref. \onlinecite{min_soi_2006}. 

\subsection{Magnetic fields}

The Zeeman Hamiltonian describing the interaction of the electron spin with the magnetic field is in the tight-binding model written as 
\begin{equation}
 H_{\rm mag}=\sum_{i, \mu, v, \lambda, \lambda'} \mu_B B_v c^\dagger_{i\mu\lambda} S^v_{\lambda\lambda'} c_{i\mu\lambda'},
\end{equation}
where the index  $\mu$ runs over all orbitals, $i$ labels the position of the atom, and $\lambda, \lambda'$ denote the spin. $B_v$ with $v = x,y,z$ are the components of the magnetic field in the global coordinate system, which we choose such that the electric field is always in $x$ direction and the CNT axis is along $\hat{\ve z}$.

The orbital effect of the magnetic field is usually expressed in terms of Peierls phases, multiplying the hopping amplitudes in the tight-binding Hamiltonian. For a magnetic field along the CNT axis, which will be of most interest in this paper, this will lead to an Aharonov-Bohm shift of the circumferential wave vector, which can be incorporated easily into the effective model, derived subsequently.

\section{\texorpdfstring{$\pi$}{pi} and \texorpdfstring{$\sigma$}{sigma} bands\label{pi_sigma_section}}
In general, all four orbitals of the second shell in CNTs are hybridized by the hopping Hamiltonian $H_{\rm hop}$. However, in the limit $R\rightarrow\infty$, the hopping amplitudes between the $p_r$ orbital and any of the orbitals $s,p_\1,p_\2$ become zero, and the hopping Hamiltonian becomes equal to the Hamiltonian of flat graphene. The $p_r$ orbitals then form the $\pi$ band and the remaining three orbitals, which are still strongly hybridized, form the $\sigma$ bands. For CNTs with finite $R$, the $\pi$ band hybridizes with the $\sigma$ bands due to the CNT curvature, which leads to non-zero hopping amplitudes $t^{p_rp_\1}_{ij},t^{p_rp_\2}_{ij},t^{p_rs}_{ij}$. However, the $\pi\sigma$ hybridization is still small for realistic CNT sizes, so that it is convenient to partition the total Hamiltonian [Eq. (\ref{full_hamiltonian})] as 
\begin{equation}
H = H_\pi + H_\sigma + H_{\pi\sigma} + H_{\sigma\pi}.
\end{equation}
The parts $H_\pi$ contain only electron operators $c_{ip_r\lambda}$ for the $p_r$ orbitals, $H_\sigma$ contains only operators for the orbitals $p_\1,p_\2,s$, and the parts $H_{\pi\sigma}$ ($H_{\sigma\pi}$) contain only the combinations $c^\dagger_{ip_r\lambda} c_{i\mu\lambda}$ ($c^\dagger_{i\mu\lambda} c_{ip_r\lambda}$ ) with $\mu=p_\1,p_\2,s$ running only over the $\sigma$ orbitals. We will use the symbols $H_\pi,H_\sigma,H_{\pi\sigma},H_{\sigma\pi}$ also for the representation of $H$ in first quantization, i.e., for complex matrices whenever this notation is more convenient.

In the following, we discuss the $\pi$-band Hamiltonian $H_\pi$ and the $\sigma$-band Hamiltonian $H_\sigma$ separately before we take into account the $\pi\sigma$ hybridization. For this, we transform to a $k$-space representation of the electron operators
\begin{equation}
c_{i\mu\lambda} = \frac1{\sqrt N} \sum_{\boldsymbol\kappa} e^{i\boldsymbol \kappa\cdot\ve R_i} c_{\boldsymbol\kappa \zeta \mu\lambda}
\end{equation}
where $N$ is the number of unit cells. The preliminary definition of the momentum $\boldsymbol \kappa= \kappa\, \hat{\ve z} + \kappa_t \hat{\ve t}$ has a momentum component $\kappa$ along the tube and one component $\kappa_t$ around the tube. Remember that the alignment of the corresponding unit vectors $\hat{\ve z},\hat{\ve t}$ relative to the lattice depends on the chirality of the tube (see Fig. \ref{fig_lattice_definitions}).

\subsection{\texorpdfstring{$\pi$}{pi} band hopping Hamiltonian\label{sect_pi_hopping}}
We start the discussion of the $\pi$ band with the limit $R\rightarrow\infty$, in which the hopping amplitude between all nearest-neighbor $p_r$ orbitals is $V^\pi_{pp}$. The transformation of the hopping terms of the $\pi$ band Hamiltonian $H_{\rm hop}^\pi$ to $k$-space gives
\begin{equation}
H_{\rm hop}^\pi = V_{pp}^\pi \sum_{\boldsymbol \kappa} w(\boldsymbol\kappa) c^\dagger_{\boldsymbol\kappa A p_r\lambda} c_{\boldsymbol\kappa B p_r\lambda} + {\rm H.c.},\label{hop_flat_pi_hamiltonian}
\end{equation}
where $w(\boldsymbol\kappa)=  e^{i\boldsymbol\kappa\cdot\ve L^1} + e^{i\boldsymbol\kappa\cdot\ve L^2} + e^{i\boldsymbol\kappa\cdot\ve L^3}$. The spectrum $\pm V_{pp}^\pi|w(\boldsymbol\kappa)|$ of $H^\pi_{\rm hop}$ is zero at the two Dirac points $\ve K=4\pi(\hat{\ve t}\cos\theta+\hat{\ve z}\sin\theta)/3a$ and $\ve K'=-\ve K$. Since we are interested mainly in the low-energy states, we expand $\boldsymbol\kappa$ about each of the two Dirac points,
\begin{equation}
\boldsymbol\kappa = \ve K + \ve k
\quad\text{or}\quad
\boldsymbol\kappa = \ve K' + \ve k,
\end{equation}
to linear order in $\ve k=k\,\hat{\ve z} + k_t\hat{\ve t}$. The resulting approximated hopping Hamiltonian for the $\pi$ band reads
\begin{equation}
H^\pi_{\rm hop} \simeq \hbar v_F \sum_{\ve k} e^{i\tau\theta} (\tau k_t - i k) c^\dagger_{\ve kA\tau p_r\lambda} c_{\ve kB\tau p_r\lambda} + {\rm H.c.},\label{pi_band_hamiltonian_raw}
\end{equation}
with $\hbar v_F = \sqrt3 |V_{pp}^\pi|/2 a $. The index $\tau=\pm1$ labels the two valleys $\ve K$ and $\ve K'$, respectively. Finally, to bring (\ref{pi_band_hamiltonian_raw}) to a more convenient form, we change the phase of all $p_r$ orbitals on the A sublattice by
\begin{equation}
c_{\ve kA\tau p_r \lambda} \rightarrow \tau e^{i\tau\theta} c_{\ve kA\tau p_r \lambda}
\end{equation}
and arrive at the usual first quantized form of the Dirac Hamiltonian
\begin{equation}
H^\pi_{\rm hop} = \hbar v_F (k_t \sigma_1 +  k \tau \sigma_2),
\label{ham_pi_hop}
\end{equation}
where the Pauli matrices $\sigma_i$ operate in the A,B sublattice space. $\sigma_3$ equals $1$ on the A sublattice and $-1$ on the B sublattice.
Note that, due to the finite circumference of carbon nanotubes, $k_t=(n-\tau\delta/3)/R$ is quantized. $\delta=(N_1-N_2)$ mod 3, for a $(N_1,N_2)$-CNT, is zero for metallic nanotubes, to which we restrict the discussion in this paper.
\begin{figure}[htp]
    \centering
    \includegraphics[width=7cm]{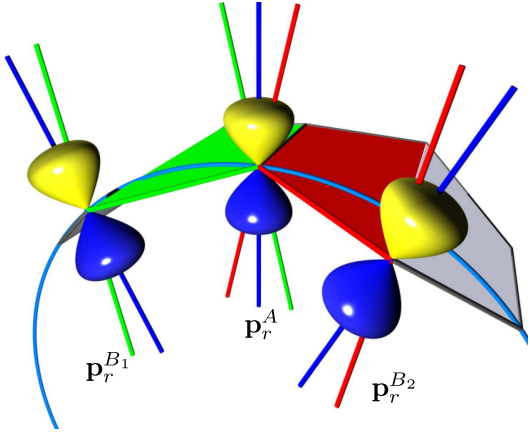}
	\caption{(Color online) Curvature induced $\sigma$ bonding between $p_r$ orbitals of neighbouring atoms
(The corresponding atoms are defined as A, B$_1$, and B$_2$). Blue lines are aligned with the corresponding $p_r$ orbital directions and red (green) lines are perpendicular to the red (green) hexagon segment and, therefore, to the red (green) bond between the atoms A and B$_2$(B$_1$). Note that all these lines (blue, red, and green) are aligned in graphene. Thus, in graphene, only $\pi$ bonding (that is due to overlap of orbitals aligned  perpendicular to the connecting line between the two atoms) is possible between such orbitals. NT curvature breaks such a symmetry: in curved graphene sheet, $p_r$ orbital direction (shown by blue lines) has perpendicular and parallel components to the line connecting two atoms. The latter component is responsible to a $\sigma$ bond formation (that is due to overlap of orbitals along the line connecting the two atoms).}
\label{curved_orbitals}
\end{figure}

The deviations of $t^{p_rp_r}_{ij}$ from $V^\pi_{pp}$ due to the finite radius $R$ leads to an additional contribution to the non-diagonal part of $H_\pi$, which is added directly to $t=V_{pp}^{\pi}w(\mathbf{k})$,
\begin{equation}
\frac{\tau a^2}{64 R^2} (4 (V^{\sigma}_{pp}+V^{\pi}_{pp})e^{-3 i \tau \theta}+e^{3 i \tau \theta}(V^{\pi}_{pp}-V^{\sigma}_{pp})).
\label{eq:h_ab_pi}
\end{equation}
This results in a shift of the wave vectors dependent on the chirality and radius along the both axes. The curvature effects also lead to a renormalization of the Fermi velocity, which can be obtained if we keep terms up to second order in $a/R$ and first order in $k$. The correction to $H^{\pi}_{\text{hop}}$ is
\begin{equation}
\Delta H_{\rm hop}^\pi =\hbar(\Delta v_F^{\perp}  k_t \sigma_1 +\tau \Delta v_F^{\parallel}  k \sigma_2),
\label{Fermi_pi}
\end{equation}
with
\begin{align}
\hbar\Delta v_F^{\perp}&= \frac{2(7V^{\pi}_{pp}+5V^{\sigma}_{pp})+\cos 6 \theta (V^{\pi}_{pp}-V^{\sigma}_{pp})}{128\sqrt{3}}\left(\frac{a}{R}\right)^2 \\
\hbar \Delta v_F^{\parallel}&= \frac{2 (V^{\sigma}_{pp} +3 V^{\pi}_{pp})-\cos 6 \theta (V^{\pi}_{pp}-V^{\sigma}_{pp})}{128\sqrt{3}}\left(\frac{a}{R}\right)^2 .
\label{renormalization_velocity}
\end{align}
Moreover, there is an additional term which couples transverse and longitudinal movements,
\begin{equation}
\Delta H_{\rm t-z}^\pi =\hbar \Delta v_F^{t-z}(( k \sigma_1 +\tau  k_t \sigma_2),
\label{Fermi_pi_kz}
\end{equation}
with
\begin{equation}
\hbar \Delta v_F^{t-z} = \frac{1}{128\sqrt{3}}\left(\frac{a}{R}\right)^2 \sin 6 \theta (V^{\pi}_{pp}-V^{\sigma}_{pp}).
\end{equation}
It should be emphasized that the corrections to the Hamiltonian given by Eqs. (\ref{eq:h_ab_pi})--(\ref{Fermi_pi_kz}) are not the only terms responsible for the shifts of the wave vectors and the Fermi velocity renormalization. Further terms leading to effects on the same order of magnitude indeed arise from the $\pi$-$\sigma$ band coupling. Such corrections by the higher energy bands will be discussed in the next sections. Collecting all terms we obtain the correction to the Fermi velocity
\begin{equation}
\frac{\Delta v_F^{\parallel}}{ v_F} =\gamma \left(\frac{a}{R}\right)^2.
\end{equation}
From Eq. (\ref{renormalization_velocity}) we find $\gamma=-0.05$, 
while the direct numerical solution of Eq. \eqref{full_hamiltonian}
gives $\gamma=-0.15$, which indeed reflects the significance of the 
correction by the energetically higher bands. In any case, the curvature-induced renormalization of the Fermi velocity is small and does not give rise to any significant effect in the regime discussed in this work.

\subsection{\texorpdfstring{$\sigma$}{sigma} band hopping Hamiltonian}

The orbitals $s,p_\1,p_\2$ on the sublattices A and B give rise to six $\sigma$ bands. For a flat honeycomb lattice (i.e., $R\rightarrow\infty$), there is no hybridization between these $\sigma$ bands and the $\pi$ band. For CNTs with finite radii $R$, finite hopping matrix elements between the $\sigma$ and the $\pi$ orbitals cause a small hybridization which will be treated in perturbation theory below. Indeed, one of the small parameters appearing in the perturbation theory is the ratio $a/R$ with $a$ the lattice constant. It is important to understand that, in order to correctly account for the curvature effects to order $(a/R)^2$ in the analytical theory derived in Sec. \ref{sect_effective_hamiltonian}, it is sufficient to treat the $\sigma$ bands in the limit $R\rightarrow\infty$, i.e., in zeroth order in $a/R$. Any curvature effect in the $\sigma$ bands of order of $a/R$ would lead to a correction of at least third order. 

Furthermore, since the effective theory will be valid only near the Dirac points $\ve K, \ve K'$, we only consider the eigenstates and eigenenergies of the $\sigma$ bands at the Dirac points, i.e., for $\boldsymbol\kappa=\ve K, \ve K'$. In $k$ space and in the basis
\begin{equation}
\mathcal{S}=\left\{ |s^A\rangle,\ |p_\1^A\rangle,\ |p_\2^A\rangle,\ |s^B\rangle,\ |p_\1^B\rangle,\ |p_\2^B\rangle\right\},
\end{equation}
the $\sigma$ band Hamiltonian has the block form
\begin{equation}
H_\sigma = \begin{pmatrix}
\ve 0 & h_{\sigma,AB} \\
h_{\sigma,AB}^\dagger &\ve 0
\end{pmatrix},
\end{equation}
where $h_{\sigma,AB} $ is given by
\begin{equation}
 \frac{3}{4} \left( \begin{matrix}
0 & 2 i V_{sp} &   2 \tau  V_{sp} \\
-2 i V_{sp}& \tau e^{ -3 i \tau \theta}(V_{pp}^\pi-V_{pp}^\sigma) & i e^{ -3 i \tau \theta}(V_{pp}^\pi-V_{pp}^\sigma)  \\
- 2 \tau V_{sp}& i e^{ -3 i \tau \theta}(V_{pp}^\pi-V_{pp}^\sigma) & - \tau e^{ -3 i \tau \theta} (V_{pp}^\pi-V_{pp}^\sigma) \end{matrix} \right).
\end{equation}
$\tau=\pm1$ labels the Dirac point $\ve K,\ve K'$, respectively. $H_\sigma$ can be diagonalized analytically. We find the energy spectrum of the $\sigma$ bands at $\ve K,\ve K'$
\begin{align}
\varepsilon^{\sigma(*)}_{pp} &= \pm \frac{3}{2}(V_{pp}^\pi-V_{pp}^\sigma),\label{eq:HsigmaK1} \\
\varepsilon^{\sigma(*)}_{sp1}&= \frac{1}{2}\left( \varepsilon_{s}\pm\sqrt{(\varepsilon_{s})^2+18(V_{sp})^2} \right),\label{eq:HsigmaK2}\\
\varepsilon^{\sigma(*)}_{sp2}&=\varepsilon^{\sigma(*)}_{sp1}.\label{eq:HsigmaK3}
\end{align}
The corresponding eigenvectors are given by
\begin{eqnarray}
|\sigma_{pp}\rangle&=&\frac{1}{\sqrt{2}}( -\tau e^{-3 i \tau \theta}|l^{A,-}\rangle + |l^{B,+}\rangle ),\label{sigma_pp}\\
|\sigma_{pp}^*\rangle&=&\frac{1}{\sqrt{2}}( \tau e^{-3 i \tau \theta}|l^{A,-}\rangle + |l^{B,+}\rangle ),\\
|\sigma_{sp1}\rangle&=& - \tau \eta_- |s^{A}\rangle + \eta_+  |l^{B,-}\rangle,\\
|\sigma_{sp2}\rangle&=&  \eta_-  |s^{B}\rangle + \tau \eta_+ |l^{A,+}\rangle,\\
|\sigma_{sp1}^*\rangle&=& \tau \eta_+ |s^{A}\rangle + \eta_-  |l^{B,-}\rangle,\\
|\sigma_{sp2}^*\rangle&=& \eta_+  |s^{B}\rangle - \tau \eta_- |l^{A,+}\rangle,\label{sigma_sp2}
\end{eqnarray}
where
\begin{eqnarray}
&|l^{j,+}\rangle=\frac{1}{\sqrt{2}} (|p_z^j\rangle + i  \tau  |p_t^j\rangle),\\
&|l^{j,-}\rangle=\frac{1}{\sqrt{2}} (|p_z^j\rangle - i \tau  |p_t^j\rangle),\\
&\eta_\pm=\frac{1}{\sqrt{2}}\sqrt{1\pm\frac{\varepsilon_{s}}{\sqrt{(\varepsilon_{s})^2+18(V_{sp})^2}}}.
\end{eqnarray}

\subsection{\texorpdfstring{$\pi\sigma$}{pisigma} hybridization}
In this section we consider different mechanisms which lead to a coupling between $\pi$ and $\sigma$ bands. All these couplings are weak compared to the typical energy scales of the $\pi$ and $\sigma$ bands, so that we may treat them in perturbation theory. It is therefore convenient to write the Hamiltonian of a carbon nanotube in first quantization in the block form
\begin{equation}
H=\begin{pmatrix}
H_{\pi} & H_{\pi\sigma}\\
H_{\sigma\pi} & H_{\sigma}
\end{pmatrix},
\label{eq:H_graphene}
\end{equation}
where $H_{\pi\sigma}$ is a $2\times6$ matrix describing the $\pi\sigma$ mixing. It contains three types of contributions, namely from the spin-orbit interaction, from the curvature of the nanotube and from the external electric field, applied to the nanotube. As we want to describe the spin-orbit coupling, we need to explicitly take into account the spin of the electron. Thus, each of the 12 matrix elements of $H_{\pi\sigma}$ is an operator acting on the electronic spin. In the matrices describing curvature and electric field effects, this operator will be the identity, but for the spin-orbit interaction matrices, it will be composed of the spin Pauli matrices $S^\mu$. Also, the matrix elements may contain momentum and position operators.

In the basis $\mathcal{P\times S}$ with $\mathcal{P}=\left\{ |p_r^A\rangle, \ |p_r^B\rangle \right\}$ and $\mathcal{S}=\left\{ |s^A\rangle,\ |p_t^A\rangle,\ |p_z^A\rangle,  |s^B\rangle,\ |p_t^B\rangle,\ |p_z^B\rangle\right\}$ the total Hamiltonian $H$ is of the form
\renewcommand{\fboxsep}{1pt}
\begin{equation}
\left(\makebox{
\begin{tabular}{cc}
		$\boxed{\begin{array}{cc}
				0\vphantom{V_{p_rp_z}^{AB}}&t\\
				t^*&0\vphantom{V_{p_rp_z}^{AB}}	
\end{array}}$
	&\renewcommand{\fboxrule}{2pt}

	$\boxed{\begin{array}{cccccc}
			V_{p_rs}^{AA}&V_{p_rp_t}^{AA}&V_{p_rp_z}^{AA}&\phantom{-}\makebox[0mm][l]{$V_{p_rs}^{AB}$}\phantom{H^{AB}_{sp_z}}&\makebox[0mm][l]{$V_{p_rp_t}^{AB}$}\phantom{H^{AB}_{p_tp_t}}&\makebox[0mm][l]{$V_{p_rp_z}^{AB}$}\phantom{H^{AB}_{p_tp_t}}\\
			V_{p_rs}^{BA}&V_{p_rp_t}^{BA}&V_{p_rp_z}^{BA}&\phantom{-}\makebox[0mm][l]{$V_{p_rs}^{BB}$}\phantom{H^{AB}_{sp_z}}&\makebox[0mm][l]{$V_{p_rp_t}^{BB}$}\phantom{H^{AB}_{p_tp_t}}&\makebox[0mm][l]{$V_{p_rp_z}^{BB}$}\phantom{H^{AB}_{p_tp_t}}\\
	\end{array}}$
	\\
	\\[-3mm]
	\raisebox{-0pt}{\renewcommand{\fboxrule}{2pt}$\boxed{\begin{array}{c}
			\makebox[1.6em]{\dotfill$\vphantom{H^{AB}_{p_tp_t}}$}\\
			\makebox[1.6em]{\dotfill$\vphantom{H^{AB}_{p_tp_t}}$}\\
			\makebox[1.6em]{\dotfill$\vphantom{H^{AB}_{p_tp_t}}$}\\[3mm]
			\makebox[1.6em]{\dotfill$\vphantom{\varepsilon_{2s}}$}\\
			\makebox[1.6em]{\dotfill$\vphantom{H^{AB}_{p_tp_t}}$}\\
			\makebox[1.6em]{\dotfill$\vphantom{H^{AB}_{p_tp_t}}$}\\
	\end{array}}$}&\hspace{-0.7pt}\raisebox{-0pt}{
	$\boxed{\begin{array}{llllll}
			\makebox[0mm][l]{$\varepsilon_{s}$}\phantom{V_{p_rs}^{AA}}&\makebox[0mm][l]{0}\phantom{V_{p_rp_t}^{AA}}&\makebox[0mm][l]{0}\phantom{V_{p_rp_t}^{AA}}&\phantom{-}H^{AB}_{ss}&H^{AB}_{sp_t}&H^{AB}_{sp_z}\\
\hdotsfor{1}&0&0&-H^{AB}_{sp_t}&H^{AB}_{p_tp_t}&H^{AB}_{p_tp_z}\\
\hdotsfor{2}&0&-H^{AB}_{sp_z}&H^{AB}_{p_tp_z}&H^{AB}_{p_zp_z}\\[3mm]
\hdotsfor{3}&\phantom{-}\varepsilon_{s}&0&0\\
\hdotsfor{4}&0&0\\
\hdotsfor{5}&0\\
	\end{array}}$}\\
\end{tabular}
}\right),
\label{eq:H_graphene_full}
\end{equation}
where, as explained above, each matrix element is an operator acting on the electron spin and, furthermore, each matrix element may depend on the crystal momentum $\boldsymbol\kappa$ or on the spatial position $\ve r$. It will turn out that the only spatial dependencies we need to deal with are dependencies on the azimuthal angle $\varphi_i$ of the carbon atoms.

The diagonal blocks correspond to the Hamiltonians of the isolated $\pi$ and $\sigma$ bands , as discussed in the two preceding subsections, and the off-diagonal blocks (framed in bold boxes) are the $H_{\pi\sigma}$ and $H_{\sigma\pi} = H_{\pi\sigma}^\dagger$ matrices with entries $V^{ij}_{uv}=\langle u^i|H_\mathrm{hyb}|v^j\rangle$ for $i,j\in \{A,B\}$ and $u,v= \in \left\{ s,p_t,p_z,p_r \right\}$). Here $H_\mathrm{hyb}$ is the part of the Hamiltonian inducing transitions between the $\pi$ and $\sigma$ bands. As stated above, it is composed of three different contributions: the hybridization coming from the curvature $H_{\pi\sigma}^{\rm curv}$, from the spin-orbit $H_{\pi\sigma}^{\rm SO}$, and from an applied electric field $H_{\pi\sigma}^{E}$. Each of these contributions will be discussed in the following.

\subsubsection{Curvature induced  \texorpdfstring{$\pi$}{pi}-\texorpdfstring{$\sigma$}{sigma} bond hybridization in nanotubes}

Due to the curvature of the nanotube surface, hoppings between orbitals on neighboring carbon atoms, which are symmetry-forbidden in flat graphene, become allowed in CNTs. This is illustrated in Fig. \ref{curved_orbitals}. We now calculate the corresponding hopping matrix elements in $k$-space at the Dirac points ($\boldsymbol\kappa=\ve K,\ve K'$) between the $\pi$ orbital ($p_r$) and the $\sigma$ orbitals ($s,p_t,p_z$). As the hopping is spin-independent, the hopping matrix is the identity in spin space. Furthermore, since the hopping does not depend on the spatial position of the electron but only on its momentum, the matrix elements will generally be functions of the momentum $\boldsymbol\kappa$ but not of the position operator. Since we are interested only in the physics near the Dirac point, it will be sufficient to consider the Dirac momenta $\boldsymbol\kappa=\ve K,\ve K'$ only. The neglect of the momentum deviation $\ve k = \boldsymbol\kappa - \ve K^{(\prime)}$ from the Dirac momenta in the $\pi\sigma$ coupling matrices turns out to be a good approximation; it only gives rise to a small renormalization of the Fermi velocity. $\ve k$ is only important in the $\mathcal P$ subspace, as discussed in Sec. \ref{sect_pi_hopping}.

It is important to note that, for a curved surface, the vectors $\ve L^n$ connecting nearest neighbors, not only have tangential components proportional to $\hat{\ve z}$ and $\hat{\ve t}$, but also a radial contribution $L_r^n \hat{\ve r}$. The magnitude $L_r^n$ is of order $a/R$ and thus vanishes in the graphene limit $R\rightarrow\infty$. The hopping amplitudes $t^{ij}_{\mu\nu}$ must be calculated as described in Eqs. (\ref{eq:hab1})--(\ref{eq:hab3}), and using the above described $\ve L^n$ vectors with radial contributions. For the transformation to $k$-space, on the other hand, the radial contributions of $\ve L^n$ are not needed because $\boldsymbol \kappa$ and $\ve k$ are defined in the two-dimensional tangent space, describing the longitudinal ($\hat{\ve z}$) and the circumferential or transverse ($\hat{\ve t}$) direction. Nevertheless, it is convenient to write scalar products between the three-component vectors $\ve L^n = L^n_z \hat{\ve z} + L^n_t \hat{\ve t} + L^n_r \hat{\ve r}$ and the momentum $\boldsymbol \kappa = \kappa \hat{\ve z} + \kappa_t \hat{\ve t} + 0 \hat{\ve r}$.

For a given $\boldsymbol\kappa$, the curvature-induced $\pi\sigma$ coupling is defined by
\begin{multline}
H^{\rm curv}_{\pi\sigma} = \sum_{\substack{n=1,2,3\\\mu,\lambda}} \biggl[e^{i\boldsymbol\kappa \cdot \ve L^n} t^{A B_n}_{p_r\mu} c^\dagger_{\boldsymbol\kappa A p_r\lambda} c_{\boldsymbol\kappa B\mu\lambda} \\+ e^{-i\boldsymbol\kappa\cdot\ve L^n} t^{B_nA}_{ p_r \mu} c^\dagger_{\boldsymbol\kappa Bp_r\lambda}c_{\boldsymbol\kappa A \mu\lambda} \biggr],
\end{multline}
where $\mu = s, p_t, p_z$ runs over the orbitals forming the $\sigma$ band. As we are finally interested in a low-energy theory for the $\pi$ band, taking the coupling to the $\sigma$ band into accound in up to second order in the small parameters, one of which is $a/R$, and since $H^{\rm curv}_{\pi\sigma}$ will only enter in second order perturbation theory, it is sufficient to keep only the linear $a/R$ order in $H^{\rm curv}_{\pi\sigma}$. Doing so, we find in the $\mathcal{P}\times \mathcal{S}$ basis
\begin{equation}
H_{\sigma\pi}^{\rm curv}=
\frac{\sqrt{3}a}{16 R} \begin{pmatrix}
                        0 & -2 \tau e^{-3 i \tau \theta }V_{sp}\\
0& i (3V^{\sigma}_{pp}+5V^{\pi}_{pp})\\
0& -\tau (V^{\pi}_{pp}-V^{\sigma}_{pp})\\
 -2 \tau e^{3 i \tau \theta }V_{sp}&0\\
i (3V^{\sigma}_{pp}+5V^{\pi}_{pp})&0\\
\tau (V^{\pi}_{pp}-V^{\sigma}_{pp})&0
                       \end{pmatrix}.\label{h_curv_pisigma_matrix}
\end{equation}
Again, $\tau=\pm1$ labels the Dirac points $\boldsymbol\kappa=\ve K,\ve K'$, respectively. Furthermore, note that $H_{\sigma\pi}^{\rm curv}$ is the identity in spin space, so that the matrix in Eq. (\ref{h_curv_pisigma_matrix}) enters the block Hamiltonian twice if the electron spin is taken into account.

\subsubsection{Spin-orbit coupling in \texorpdfstring{$\pi$}{pi} bands}

For the analysis of the spin-orbit interaction it is important to note that the corresponding Hamiltonian $H_{\rm SO}$ [Eq. (\ref{hamiltonian_spin_orbit})] is local, i.e., it has no matrix elements connecting orbitals from different lattice sites. Thus, $H^{\rm SO}_{\sigma\pi}$ can be represented as a $6\times 2\times 2$ tensor, corresponding to the six $\sigma$ orbitals per unit cell, the two $\pi$ orbitals per unit cell and the two-dimensional spin space. Each matrix element, however, may depend on the unit cell coordinate. It is most convenient to write $H^{\rm SO}_{\sigma\pi}$ in a local spin basis with spin-components $S^r$ pointing in the radial direction of the nanotube, $S^t$ pointing in the transverse (or circumferential) direction, and $S^z$ pointing along the tube axis. Since the local environment, i.e., the definition of the direction of the $p$ orbitals, are equal for all lattice sites of the tube, the $\pi\sigma$ coupling matrix due to spin-orbit interaction has a simple form
\begin{equation}
H_{\sigma\pi}^{SO}=i\Delta_{SO}\begin{pmatrix}
0&0\\
 S^z&0\\
- S^t &0\\
0&0\\
0& S^z\\
0&- S^t
\end{pmatrix}.\label{soi_pi_sigma_coupling_matrix}
\end{equation}
Note that there are no $S^r$ operators in the matrix elements of the SOI 
between the $p_r$ and the $\sigma$ orbitals. This is easily seen from the second quantized form of the atomic spin-orbit Hamiltonian Eq. (\ref{hamiltonian_spin_orbit}), in which the Levi Civita symbol forbids the terms in which a spin operator and an orbital in the same direction appear. As a $\pi\sigma$ coupling always involves a $p_r$ orbital, the $S^r$ spin cannot appear.

In Eq. (\ref{soi_pi_sigma_coupling_matrix}) the directions of the spin operators are defined in the local basis $\hat{\ve r}, \hat{\ve t}, \hat{\ve z}$. On the other hand, the hopping Hamiltonian does not affect the electron spin. A electron with spin pointing in the global $x$ direction, say, and is hopping around the nanotube has its spin pointing in the global $x$ direction independently of its position. This means that the hopping Hamiltonian is the identity in spin space, but only if the global spin basis $S^{x,y,z}$ is used, rather than the local spin basis $S^{r,t,z}$.  Thus, also the spin operators in Eq. (\ref{soi_pi_sigma_coupling_matrix}) must be transformed to the global basis. This transformation is given by
\begin{align}
S^r(\varphi_i) &= S^x \cos\varphi_i + S^y \sin\varphi_i \\
S^t(\varphi_i) &= S^y \cos\varphi_i - S^x \sin\varphi_i \\
S^z &= S^z.
\end{align}

Note that, unlike the hopping matrix $H_{\sigma\pi}^{\rm curv}$, $H^{\rm SO}_{\sigma\pi}$ depends not on the momentum but rather on $\varphi_i$. This is because the spin-orbit Hamiltonian we started from did not involve hoppings between neighboring carbon atoms but only local terms. Transforming this spatially dependent part of the Hamiltonian to the momentum space leads to non-diagonal matrix elements in $\ve k$, coupling the transverse momentum $k_t$ to its neighboring momenta $k_t\pm \frac1R$. As we are finally interested in a low-energy theory for the lowest $\pi$ subband, all virtual $+\Delta k_t$ processes must be compensated by a $-\Delta k_t$ process in second order perturbation theory. This will be discussed in detail in Sec. \ref{sect_effective_hamiltonian}. For now, we keep the real-space notation of $H_{\sigma\pi}^{\rm SO}$ and emphasize again that the matrices in Eqs. (\ref{h_curv_pisigma_matrix}) and (\ref{soi_pi_sigma_coupling_matrix}) are defined with respect to a different basis ($k$ space and real space, respectively).

\subsubsection{Electric field}
As discussed in Sec. \ref{sec_definitions_el_field}, there are two significant effects of an electric field applied perpendicular to the carbon nanotube. The orbitally diagonal cosine potential, described by $H^{(1)}_E$, will be discussed in Sec. \ref{section_screening}; it turns out that $H^{(1)}_E$ is reduced by screening effects and, apart from leading to a small renormalization of the Fermi velocity,\cite{} has no significant effect on the low-energy theory we aim at.

Most important, however, is the $s$-$p_r$ transition, described by $H^{(2)}_E$. In the basis $\mathcal{P\times S}$, we find for the contribution of the $s$-$p_r$ transition to the $\pi\sigma$ coupling
\begin{equation}
H_{\sigma\pi}^{E}=\begin{pmatrix}
-e E_r (\varphi_i)\xi_0 & 0\\
0&0\\
0&0\\
0& -e E_r(\varphi_i) \xi_0\\
0&0\\
0&0
\end{pmatrix},
\end{equation}
where $E_r (\varphi_i)$ is the radial component of the electric field. The angular dependence of the electric field in the linear response regime can be approximated by
\begin{equation}
E_r(\varphi_i)= E \cos\varphi_i,
\end{equation}
where $E$ is the magnitude of the applied electric field applied perpendicular to the CNT axis (see also Fig. 1 in Ref. \onlinecite{klinovaja_helical_modes_2011}). Again, we note that the matrix elements of $H_{\sigma\pi}^{E}$ are identities in spin space and do not depend on the momentum but on the azimuthal angle $\varphi_i$ of the carbon atoms.

\section{Effective Hamiltonian for the \texorpdfstring{$\pi$}{pi} band\label{sect_effective_hamiltonian}}

Having defined all parts of the microscopic Hamiltonian, we are now in a position to derive the effective low-energy Hamiltonian of the CNT. 
This will be done in second order perturbation theory. The small parameters in which we expand are: the surface curvature $a/R$, the spin-orbit interaction coupling strength $\Delta_{\rm SO}=6$  meV, and the electric field strength $eE\xi_0 \simeq 50\cdot E[{\rm V/nm}]$ meV. The last two quantities are energy scales and must be compared to the typical $\sigma$ band eigen energies which are on the order of a few eV. In the derivation of the low-energy theory, we completely neglect the Hamiltonian $H_E^{(1)}$. For the $\pi$ band this Hamiltonian alone is known to give rise to a Fermi-velocity renormalization of second order in the electric field.\cite{novikov_vF_renormalization_2006} The question whether $H_E^{(1)}$ can actually be neglected will be critically discussed in Sec. \ref{sect_analysis_eff_theory}.

The effective Hamiltonian for the $\pi$ band is calculated in second-order perturbation theory as
\begin{eqnarray}
H_{\pi}^\mathrm{eff}&\simeq&H_{\pi}+H_{\pi\sigma}\frac{1}{\varepsilon-H_\sigma}H_{\pi\sigma}^\dagger, \label{eq:foldingdown},
\end{eqnarray}
where
\begin{eqnarray}
H_{\pi\sigma}&=&H_{\pi\sigma}^{\rm curv}+H_{\pi\sigma}^{SO} +H_{\pi\sigma}^{E}.
\end{eqnarray}
We proceed by inserting into Eq. \eqref{eq:foldingdown} the unitary matrix $U_{\sigma}$, which  diagonalizes $H_\sigma$ and is constructed from the eigenvectors given by Eqs. (\ref{sigma_pp})--(\ref{sigma_sp2})
\begin{eqnarray}
H_{\pi}^\mathrm{eff}&\simeq&H_{\pi}+H_{\pi\sigma}U_{\sigma}U^\dagger_{\sigma}\frac{1}{\varepsilon-H_\sigma}U_{\sigma}U^\dagger_{\sigma}H_{\pi\sigma}^\dagger,
\end{eqnarray}
so that the inverse operator $(\varepsilon-H_\sigma)^{-1}$ reduces to a diagonal sum of the $\sigma$ band eigenvalues. Furthermore, since we are interested in energies close to the Dirac point, we set $\varepsilon=0$. This reduces the complicated $8\times 8$ Hamiltonian matrix, describing $\pi$ and $\sigma$ orbitals, to an effective $2\times 2$ Hamiltonian matrix for the $\pi$ orbitals only. The two-dimensional vector space, the matrix $H^{\rm eff}_\pi$ is defined in, corresponds to the two sublattices A and B. Note, however, that the matrix elements of $H^{\rm eff}_\pi$ still contain the momentum operator $\hat{\ve k} = (\hat k_t,\hat k)$, the position operator $\hat \varphi_i$ and spin operators. In particular, we find
\begin{multline}
H_{\pi}^\mathrm{eff}(\hat\varphi_i,\hat {\ve k}) = \hbar v_F \bigl[ (\hat k_t+\Delta k_{\rm cv}^t)\sigma_1 +\tau ( \hat k+ \Delta k_{\rm cv}^z  ) \sigma_2\bigr] \\ +\alpha\bigl[S^z\sigma_1- \tau S^t(\varphi_i)\sigma_2\bigl]
+\tau \beta_1 \bigl[ S^z \cos 3 \theta - S^t(\varphi_i)\sin 3 \theta  \bigr] \\ + \tau e E_r  (\varphi_i) \xi_2  \bigl[S^t(\varphi_i)\sigma_2 - \tau S^z\sigma_1\bigr]+\\+ \gamma_1 \tau S^r(\varphi_i) \sigma_3 + e E_r (\varphi_i) \xi_1,
\label{h_pi_angle_dependent}
\end{multline}
with the coefficients 
\begin{align}
\hbar v_F \begin{pmatrix}\Delta k_{\rm cv}^t\\\Delta k_{\rm cv}^z\end{pmatrix} &=\tau \dfrac{V^{\pi}_{pp}(V^{\pi}_{pp}+V^{\sigma}_{pp})}{8 (V^{\pi}_{pp}-V^{\sigma}_{pp})}\left(\dfrac{a}{R}\right)^2 \begin{pmatrix} -\cos 3\theta\\\sin 3\theta\end{pmatrix}\\
\alpha &= \frac{\sqrt{3}\varepsilon_s(V_{pp}^\pi+V_{pp}^\sigma)\Delta_{\rm SO}}{18 V_{sp}^2} \frac{a}{R},\\
\beta_1 &= -\frac{\sqrt{3} V_{pp}^\pi \Delta_{\rm SO}}{ 3 (V_{pp}^\pi-V_{pp}^\sigma)}\frac{a}{R},\\
\gamma_1 &= \frac{2 \varepsilon_s \Delta_{\rm SO}^2 }{9 V_{sp}^2},\\
\xi_1 &=  -\frac{(V_{pp}^\sigma+V_{pp}^\pi)}{2 \sqrt{3} V_{sp}}\frac{a}{R}\xi_0,\\
\xi_2 &= \frac{2 \Delta_{\rm SO}}{3 V_{sp}} \xi_0.
\end{align} 

As we aim at an effective theory for the lowest subband, we project $H^{\rm eff}_{\pi}(\hat\varphi_i,\hat{\ve k})$ onto the subspace spanned by the wave functions of this subband. These wave functions are plane waves $\propto \exp(i k z)$ which do not depend on $\varphi$. Thus, projection means $\varphi$-averaging the Hamiltonian and setting $\hat k_t =0$. Furthermore, since there is no operator in $H_{\pi}^\mathrm{eff}(\hat\varphi_i,\hat {\ve k})$ which does not commute with $\hat k$, we may write
\begin{equation}
H_\pi^\mathrm{eff}(k) = \int  \frac{d\varphi}{2\pi} H_\pi^\mathrm{eff} (\varphi,(0,k)).
\end{equation}
Any term in $H_{\pi}^\mathrm{eff}(\varphi,(0,k))$ containing odd powers of sin or cos average to zero. Finally, the resulting effective low-energy theory is given by
\begin{eqnarray}
H_\pi^{\rm eff} & = & H^\pi_{\rm hop} + H_{\rm orb}^{\rm cv} + H_{\rm SO}^{\rm cv} + H_{\rm SO}^{\rm el}.
\label{eq:effective_hamiltonian}
\end{eqnarray}

$H_{\rm orb}^{\rm cv}$ describes the curvature induced $k$-shift of the Dirac points.\cite{ando_soi_cnt_2000, izumida_soi_cnt_2009, white_soi_cnt_2009} For non-armchair metallic CNTs, $\Delta k_{\rm cv}^t \neq 0$ leads to the opening of a gap at the Dirac point
\begin{eqnarray}
H_{\rm orb}^{\rm cv} &=& \hbar v_F \Delta k_{\rm cv}^t \sigma_1,
\end{eqnarray}
with
\begin{eqnarray}
\hbar v_F\Delta k_{\rm cv}^t&=& -\tau \dfrac{5.4\, {\rm meV}}{R[{\rm nm}]^2}\cos 3\theta.
\end{eqnarray}
The $k$-shift $\Delta k_{\rm cv}^z$ along the CNT is irrelevant and has been neglected here (see also Ref. \onlinecite{klinovaja_helical_modes_2011}).

$H_{\rm SO}^{\rm cv}$ contains the curvature induced SOI \cite{jeong_soi_cnt_2009, izumida_soi_cnt_2009}   which does not average out in the $\varphi$ integration,
\begin{eqnarray}
H_{\rm SO}^{\rm cv}&=& \alpha S^z\sigma_1 + \tau \beta S^z,
\label{alpha-beta}
\end{eqnarray}
with the parameters
\begin{eqnarray}
\alpha &\simeq&\dfrac{-0.08\,{\rm meV}}{R[{\rm nm}]},\\
\beta &\simeq&\dfrac{-0.31\, {\rm meV}}{R[{\rm nm}]} \cos 3 \theta.
\end{eqnarray}
This term contains only on the $S^z$ spin operator which is consistent with the rotational symmetry of the nanotube.
$H_{\rm SO}^{\rm cv}$ is responsible for the breaking of the electron-hole symmetry.\cite{kuemmeth_exp_cnt_2008}

Finally, the electric field induced SOI depends on the product $E(\varphi) S^t(\varphi)$, which does not average out in the $\varphi$ integral. Indeed, $E(\varphi) \propto \cos\varphi$ while $S^t(\varphi) = S^y\cos\varphi - S^x \sin\varphi$. The $\cos^2\varphi$ integral leads to a nonvanishing 
\begin{equation}
H_{\rm SO}^{\rm el}= \tau e E \xi S^y \sigma_2,
\label{ham_electric_field}
\end{equation}
where 
\begin{equation}
\xi =\xi_2 /2=\frac{ \Delta_{\rm SO}}{3 V_{sp}} \xi_0 \simeq 2 \times 10^{-5} {\rm nm}.
\end{equation}
This term breaks rotational invariance of the CNT and involves the $S^y$ operator for the electric field along the $x$ axis. The fact that the directions of the applied field and the induced spin polarization are perpendicular is typical for the electric field-induced SOI.

\section{Field screening\label{section_screening}}

An electric field applied perpendicularly to a carbon nanotube induces a rearrangement of the electrons on the tube surface. If the electrons were free to move, as in the case of a metal cylinder, the field inside the tube would be perfectly screened. However, since the density of states in CNTs at half-filling is small, the electrons rather behave as in half-metals, and thus the field screening is only partial. In the following, we calculate the field screening and its consequences explicitly by two methods: by a linear response calculation and by a direct diagonalization of the full Hamiltonian, including the external field. The analytical linear response calculation will provide us with a simple and intuitive picture of the screening but some uncontrolled approximations such as, for instance, the restriction to the $\pi$ band, are needed there. The numerical calculation takes into account all bands derived from the second shell orbitals of the carbon atoms and provides a quantitative result, corroborating the linear response calculation.

The task is, therefore, to calculate the charge response of a CNT to the electrostatic potential described by
\begin{equation}
H_E = \phi_{0,\rm tot} \sum_{\ve n,\zeta} \cos(\varphi_{\ve n,\zeta}) \hat n_{\ve n,\zeta}\label{ham_lin_response_ext_field}
\end{equation}
with $\phi_{0,\rm tot}$ the amplitude of the total electrostatic potential acting on the tube surface. $\varphi_{\ve n,\zeta}$ is the azimuthal angle of the carbon atom in unit cell $\ve n$ and sublattice $\zeta$. The operator $\hat n_{\ve n,\zeta} = \sum_\mu c^\dagger_{\ve n,\zeta,\mu} c_{\ve n,\zeta,\mu}$ counts the electrons on atom $(\ve n,\zeta)$ in all second shell orbitals $\mu = s,p_r,p_t,p_z$. In order to keep the notation simple, we drop the spin index $\lambda$ in this section and multiply the charge response by a factor of 2.

Note that $H_E$ describes a homogeneous electric field $E$, which gives rise to the electrostatic potential $\phi_{\rm ext}(\varphi)= eER\cos\varphi$. $E$ is the field which is applied externally. In the following we will show that the induced electron charges on the tube surface give rise to an induced electrostatic potential $\phi_{\rm ind}(\ve r)$ which, although it has a rather complicated spatial structure away from the nanotube surface (i.e. for $|\ve r|\neq R$), reduces to $\phi_{\rm ind}(\varphi) = \phi_{0,\rm ind}\cos\varphi$ for $|\ve r|=R$. Thus, anticipating this result, the total potential $\phi_{\rm tot}(\varphi)= \phi_{\rm ext}(\varphi) + \phi_{\rm ind}(\varphi) = \phi_{0,\rm tot}\cos\varphi$ is described by the Hamiltonian (\ref{ham_lin_response_ext_field}). As discussed in Sec. \ref{subsec_screening}, this additivity property of the amplitudes $\phi_{0,\rm ext} $ and $\phi_{0,\rm ind}$ can be interpreted as a linear screening of the electric field inside the nanotube, i.e., $\phi_{0,\rm tot}=eE^* R$ with the screened field $E^*<E$.

\subsection{Linear response of the \texorpdfstring{$\pi$}{pi} band}
We define the static linear response coefficient $\chi^{\mu\mu'}_{\zeta\zeta'}(\ve n,\ve n')$ as the proportionality constant between the density response at site $(\ve n,\zeta)$ in orbital $\mu$ and the density perturbation at site $(\ve n',\zeta')$ in orbital $\mu'$ described by any Hamiltonian $A c^\dagger_{\ve n'\zeta'\mu'}c_{\ve n'\zeta'\mu'}$
\begin{equation}
\rho_{\ve n\zeta\mu} = \delta\left<c^\dagger_{\ve n\zeta\mu}c_{\ve n\zeta\mu} \right> = A \chi^{\mu\mu'}_{\zeta\zeta'}(\ve n,\ve n').
\end{equation}
$\chi^{\mu\mu'}_{\zeta\zeta'}(\ve n,\ve n')$ can be calculated by the Kubo formula
\begin{multline}
\chi^{\mu\mu'}_{\zeta\zeta'}(\ve n,\ve n') = -2 i \int_0^\infty \D t \, e^{-\eta t} \\ \times \left<\left[c^\dagger_{\ve n\zeta\mu}(t) c_{\ve n\zeta\mu}(t),c^\dagger_{\ve n'\zeta'\mu'}c_{\ve n'\zeta'\mu'}\right]\right>,
\end{multline}
where $c_{\ve n,\zeta,\mu}(t) = e^{it H} c_{\ve n,\zeta,\mu} e^{-it H}$ is the Heisenberg representation of the electron annihilation operator and $\eta=0^+$ ensures the convergence of the time integral. The factor 2 accounts for the spin-degeneracy. The average denotes the expectation value with respect to the ground state of the electronic system. The generalization to finite temperatures is possible but not of interest here.

The linear response of the nanotube to the Hamiltonian defined in Eq. (\ref{ham_lin_response_ext_field}) is
\begin{align}
\rho_{\ve n\zeta\mu}&= \phi_0 \sum_{\ve n',\zeta',\mu'} \cos(\varphi_{\ve n',\zeta'}) \chi^{\mu\mu'}_{\zeta\zeta'}(\ve n,\ve n').
\end{align}
In the remainder of this subsection we deal only with the linear response in the $\pi$ band allowing us to set $\mu=\mu'=p_r$ and suppress the orbital index. The contributions of the $\sigma$ band will be discussed below on the basis of the numerical calculation. Furthermore, we restrict the calculation to armchair nanotubes.

A straightforward calculation of the charge density induced by the Hamiltonian (\ref{ham_lin_response_ext_field}) gives rise to two terms in the charge response: a normal response, which  has the same cosine modulation as the inducing Hamiltonian [Eq. (\ref{ham_lin_response_ext_field})], and an anomalous response, which is staggered on the sublattice level and has a sine modulation,
\begin{equation}
\rho(\varphi_{\ve n,\zeta}) = \phi_{0,\rm tot}  \left[\chi_n \cos\varphi_{\ve n,\zeta} - \zeta \chi_a \sin\varphi_{\ve n,\zeta}\right],\label{charge_response_form}
\end{equation}
with the normal and anomalous response coefficients
\begin{align}
\chi_n &= \chi_{AA} + {\rm Re}\chi_{BA} \cos(\varphi_{AB}) - {\rm Im}\chi_{BA} \sin(\varphi_{AB}), \\
\chi_a &= {\rm Re}\chi_{BA} \sin(\varphi_{AB}) + {\rm Im}\chi_{BA} \cos(\varphi_{AB}).
\end{align}
Here $\varphi_{AB}=a/\sqrt3 R$ is the difference of the azimuthal angle between the different sublattice sites within one unit cell and $\chi_{AA},\chi_{BA}$ are the $k$-space susceptibilities, defined by
\begin{align}
\chi_{\zeta\zeta'} &= \chi^{p_rp_r}_{\zeta\zeta'}(\ve q = \hat{\ve t}/R)\label{chiaachiab}.
\end{align}
As expected, the relevant susceptibilities for the charge response to a transverse homogeneous field are to be evaluated for zero momentum along the tube and for the smallest possible momentum around the tube.

The charge susceptibilities required for Eq. (\ref{chiaachiab}) can be expressed as $k$-space integrals
\begin{align}
\chi_{\zeta\zeta}(\ve q) &= \frac1{N} \sum_{\boldsymbol \kappa,a,a'} \frac{\Theta(\varepsilon_F - \varepsilon_a(\boldsymbol \kappa))\Theta(\varepsilon_{a'}(\boldsymbol \kappa+\ve q)-\varepsilon_F)}{\varepsilon_a(\boldsymbol \kappa) - \varepsilon_{a'}(\boldsymbol \kappa+\ve q)} \\
\chi_{-\zeta\zeta}(\ve q) &= \frac1{N} \sum_{\boldsymbol \kappa,a,a'} aa' \exp\left(i \zeta(\phi_{\boldsymbol \kappa} - \phi_{\boldsymbol \kappa+\ve q})\right) \nonumber \\ & \bs\times \frac{\Theta(\varepsilon_F - \varepsilon_a(\boldsymbol \kappa))\Theta(\varepsilon_{a'}(\boldsymbol \kappa+\ve q)-\varepsilon_F)}{\varepsilon_a(\boldsymbol \kappa) - \varepsilon_{a'}(\boldsymbol \kappa+\ve q)},
\end{align}
where the spin-degeneracy has been taken into account. $\varepsilon_F$ is the Fermi energy and $\varepsilon_a(\boldsymbol \kappa)$ are the energy eigenvalues for the two branches ($a=\pm1$) of the $\pi$ band. We assume here that the curvature effects, discussed in Sec. \ref{pi_sigma_section}, do not affect the final charge polarization significantly. This assumption will be tested numerically in the next subsection. Thus, we calculate $\varepsilon_\pm(\boldsymbol \kappa) = \pm |V_{pp}^\pi w(\boldsymbol \kappa)|$ and $\phi_{\boldsymbol \kappa} = \arg w(\boldsymbol \kappa)$ from the Hamiltonian (\ref{hop_flat_pi_hamiltonian}).

Furthermore, we restrict the discussion to the charge neutrality point $\varepsilon_F=0$ so that $a=-1$ and $a'=1$ and
\begin{align}
\chi_{\zeta\zeta}(\ve q) &= - \frac1{|V_{pp}^\pi| N} \sum_{\boldsymbol \kappa} \frac{1}{|w(\boldsymbol \kappa)|  + |w(\boldsymbol \kappa+\ve q)|} \label{tight_binding_sus_oo}\\
\chi_{-\zeta\zeta}(\ve q) &=  \frac1{|V_{pp}^\pi| N} \sum_{\boldsymbol \kappa}  \frac{\exp\left(i \zeta(\phi_{\boldsymbol \kappa} - \phi_{\boldsymbol \kappa+\ve q})\right)}{|w(\boldsymbol \kappa)| + |w(\boldsymbol \kappa+\ve q)|}\label{tight_binding_sus_moo}.
\end{align}
  Note that for $\ve q=0$ we have $\chi_{\zeta\zeta}(0) = -\chi_{-\zeta\zeta}(0)$ so that the linear response to a homogeneous potential is zero in flat graphene. This is a consequence of the vanishing density of states of graphene at the charge neutrality point. For the $\ve q$ vectors given by Eq. (\ref{chiaachiab}) and for the $k$-space grid defined by the circumference and the length of the CNT (we assume periodic boundary conditions in $z$ direction), the $k$-space summations in Eqs. (\ref{tight_binding_sus_oo}) and (\ref{tight_binding_sus_moo}) are evaluated numerically. For instance, for a (10,10)-CNT and in the limit of infinitely long tubes we obtain
\begin{align}
|V_{pp}^\pi|\chi_n &= -0.0424, & |V_{pp}^\pi| \chi_a &= 0.002.
\end{align}

\begin{figure}[!ht]
\centering
\includegraphics[width=\columnwidth]{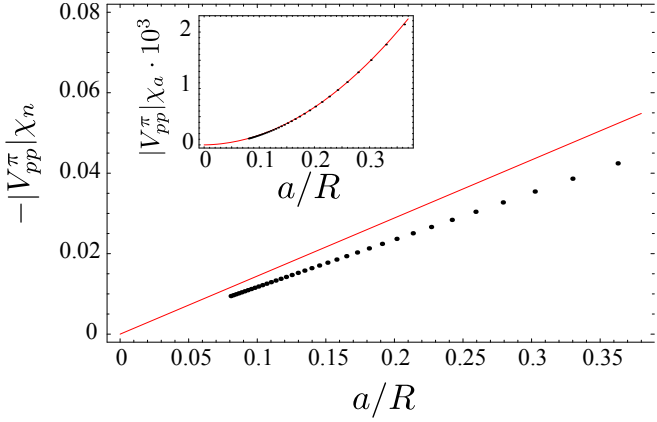}
\caption{(Color online) The normal response coefficient $\chi_n$ of the $\pi$ band as a function of $a/R$. The black dots are the exact result from the numerical evaluation of Eqs. (\ref{tight_binding_sus_oo}) and (\ref{tight_binding_sus_moo}). The solid line shows the linear term in Eq. (\ref{ana_normal_response}). The inset shows the anomalous response coefficient $\chi_a$ together with a polynomial fit $\chi_a\simeq 0.016 (a/R)^2$.}
\label{fig_screening_coefficients_kubo_and_extrapolation}
\end{figure}

In the Dirac approximation of the band structure, i.e., $\varepsilon_\pm(\ve k) = \pm \sqrt3 |V_{pp}^\pi|  |\ve k a|/2$, Eqs. (\ref{tight_binding_sus_oo}) and (\ref{tight_binding_sus_moo}) can be calculated analytically for small $q=|\ve q|$. For the spin-susceptibility, which is equal to the charge susceptibility for non-interacting systems, this has been done in Refs. \onlinecite{brey_graphene_spin_susceptibility_2007,black_schaffer_graphene_spin_susceptibility_2010,
saremi_graphene_spin_susceptibility_2007} with the result
\begin{align}
\frac{\hbar v_F}a\chi_{\zeta\zeta}(q) &= -\frac{\Lambda a
}{2\pi} + \frac{q a}{16} \label{chioo_dirac}\\
\frac{\hbar v_F}a\chi_{-\zeta\zeta}(q) &= \frac{\Lambda a
}{2\pi} - \frac{3q a}{16}\label{chimoo_dirac},
\end{align}
where $\Lambda$ is an ultraviolet cutoff, $a$ is the lattice constant, and $\hbar v_F/a=\sqrt3|V_{pp}^\pi|/2$. From this we find the analytical forms for the normal and anomalous response coefficients
\begin{align}
|V_{pp}^\pi|\chi_n &= -\frac1{4\sqrt3} \frac aR - \frac{\Lambda a}{6 \sqrt3 \pi} \left(\frac aR\right)^2 + O\left((a/R)^3\right) \label{ana_normal_response}\\
|V_{pp}^\pi|\chi_a &= \frac{\Lambda a}{3\pi}\frac aR - \frac{1}{8} \left(\frac aR\right)^2 + O\left((a/R)^3\right)\label{ana_anomalous_response}.
\end{align}
The comparison of the analytical result for $\chi_n$ with the numerical evaluation of the Kubo integrals in Fig. \ref{fig_screening_coefficients_kubo_and_extrapolation} shows that the first term in Eq. (\ref{ana_normal_response}) captures the leading $a/R$-term qualitatively in a correct way. However, the Dirac approximation leads to a wrong prefactor.

The analytical form of the anomalous response calculated in the Dirac approximation differs from the exact result. In the next subsection we explain the reason for this and why the evaluation of Kubo formulas within the Dirac approximation is not reliable. However, as we are finally interested in the field screening, the anomalous response, which is staggered on the atomic scale, is irrelevant as it averages out in a continuum approximation of the charge distribution. The staggered potential induced by the anomalous response has a $\sin\varphi$ modulation so that it does not open a gap. We have checked numerically that such an additional term in the Hamiltonian does not change our results.

\subsection{Analysis of the Kubo integral}

It is instructive to study the structure of the Kubo integrals for the charge susceptibilities [Eqs. (\ref{tight_binding_sus_oo}) and (\ref{tight_binding_sus_moo})] in more detail, especially in view of their Dirac approximations. 

For $\ve q=0$ the intra-sublattice susceptibility has the structure
\begin{equation}
\chi_{\zeta\zeta}(\ve q=0) \propto \int \D \varepsilon \frac{D(\varepsilon)}{\varepsilon},\label{chi_integral_structure}
\end{equation}
where $D(\varepsilon)=N^{-1}\sum_{\boldsymbol\kappa,a}\delta(\varepsilon-\varepsilon_a(\boldsymbol\kappa))$ is the density of states. In graphene, $D(\varepsilon)$ is known to be linear in the energy $\varepsilon$ near the Dirac point. For carbon nanotubes, which can be viewed as graphene with one confined direction, $D(\varepsilon)$ is constant for $\varepsilon\simeq0$, but it increases discontinuously with $\varepsilon$ as $\varepsilon$ crosses more and more transverse subbands. On a coarse grained energy scale $D(\varepsilon) \sim \varepsilon$ for both, CNTs and graphene, as long as $\varepsilon\lesssim 3$eV. In the Dirac approximation, it is assumed that only electronic states with low energies are important, i.e., that the integral in Eq. (\ref{chi_integral_structure}) converges before $\varepsilon\simeq 3$eV. However, Eq. (\ref{chi_integral_structure}) does obviously not converge for $D(\varepsilon)\sim \varepsilon$ so that, strictly speaking, the Dirac approximation is not allowed for Kubo formulas. 

This convergence problem is reflected in the cutoff dependence of the first terms in Eqs. (\ref{chioo_dirac}) and (\ref{chimoo_dirac}). However, the terms linear in $q$ do not depend on the ultraviolet cutoff, which suggests that only contributions from small energies, where the Dirac approximation is valid, enter the $q$-dependence. Indeed, one finds that in the high energy regime
\begin{equation}
\chi_{\zeta\zeta}(\ve q) - \chi_{\zeta\zeta}(0) \sim \int\D \varepsilon \frac{D(\varepsilon)}{\varepsilon^5}
\end{equation}
converges quickly.

In order to compare the Kubo integrals of the tight-binding formulation with the Dirac approximation, we consider $\chi_{AA}(\ve q=0)$. Eq. (\ref{tight_binding_sus_oo}) can be written as
\begin{equation}
-|V_{pp}^\pi| \chi_{\zeta\zeta}(0) = \int_0^\infty \D \varepsilon f(\varepsilon),
\end{equation}
with
\begin{equation}
f(\varepsilon) = \frac1N \sum_{\boldsymbol\kappa} \frac{\delta  (\varepsilon-|w(\boldsymbol\kappa)|)}{2|w(\boldsymbol\kappa)|}.\label{fOfE_definition}
\end{equation}
In the Dirac approximation in which $w(\ve K+\ve k) \propto k_x+ik_y$ is assumed (we may drop all the prefactors in this analysis in favor of notational simplicity),
$f_{\rm Dirac}(\varepsilon)$ can be calculated easily. Since $f(\varepsilon)\sim D(\varepsilon)/\varepsilon$, the Dirac approximation $f_{\rm Dirac}(\varepsilon)$ must be constant.

\begin{figure}[!ht]
\centering
\includegraphics[width=\columnwidth]{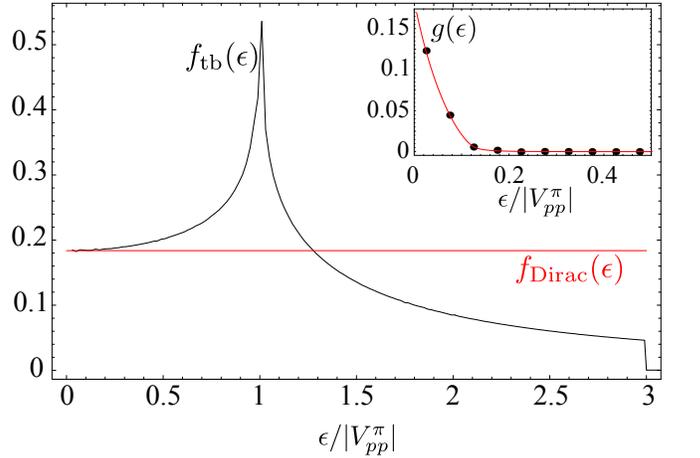}
\caption{(Color online) The energy-resolved integrand of the Kubo integral $f(\varepsilon)$ for $\chi_{\zeta\zeta}(\ve q=0)$. The black line is the integrand calculated from the full tight-binding model $f_{\rm tb}(\varepsilon)$ (see text). The horizontal gray (red) line shows the integrand in Dirac approximation $f_{\rm Dirac}(\varepsilon)$. The inset shows the integrand $g(\varepsilon)$ for $\ve q=2\pi \hat{\ve t}/50$. The dots are the results of the tight-binding calculation for a $200\times200$ $k$-space grid and the line shows the Dirac approximation of $g(\varepsilon)$.}
\label{fig_kubo_integrands}
\end{figure}

Calculating $f(\varepsilon)$ numerically for the more complicated tight-binding form of $w(\boldsymbol\kappa)$, which can be evaluated only for finite size systems $N<\infty$, is not quite straightforward. We do this by discretizing the energy in $\Delta \varepsilon$ steps and defining
\begin{multline}
f_{\rm tb}(\varepsilon,N) \\= \frac{1}{N\Delta \varepsilon}\sum_{\boldsymbol\kappa}\frac{\Theta[|w(\boldsymbol\kappa)| - \varepsilon]\Theta[\varepsilon+\Delta \varepsilon -|w(\boldsymbol\kappa)|]}{2|w(\boldsymbol\kappa)|}.\label{discrete_integrand1}
\end{multline}
In the limit $\Delta \varepsilon\rightarrow0$ and $N\rightarrow\infty$, Eq. (\ref{discrete_integrand1}) approaches the actual $f(\varepsilon)$, as defined in Eq. (\ref{fOfE_definition}). However, finite-size effects make $f_{\rm tb}$ look very rugged. In order to obtain a smooth $f(\varepsilon)$, which can be plotted nicely, we define an average over $N$
\begin{multline}
f_{\rm tb}(\varepsilon) = \frac1{2M} \sum_{m=1}^M \\ \biggl[ f_{\rm tb}(\varepsilon,(202+3m)^2) + f_{\rm tb}(\varepsilon,(203+3m)^2) \biggr],
\end{multline}
where we have assumed a quadratic $k$-space grid and we ommited all grids in which a $k$-space point hits a Dirac point. The choice of the smallest $N= 202^2$ is completely arbitrary, as well as the choice of $M=82$, as long as $M$ is large enough.

Fig. \ref{fig_kubo_integrands} shows $f_{\rm Dirac}(\varepsilon)$ and $f_{\rm tb}(\varepsilon)$. Obviously, for small $\varepsilon\ll|V_{pp}^\pi|$ the Dirac approximation coincides with the tight-binding calculation. However, for larger energies they strongly differ. Moreover, while the finite tight-binding bandwidth ($|w(\boldsymbol\kappa)|\leq3$) gives a natural high energy cutoff and ensures the convergence of the energy integral, the Dirac model must be replenished by an ultraviolet cutoff $\Lambda$ in order to ensure the convergence.

For the deviation of the finite $\ve q$ susceptibility from $\chi_{\zeta\zeta}(0)$
\begin{equation}
-|V_{pp}^\pi| [\chi_{\zeta\zeta}(\ve q) - \chi_{\zeta\zeta}(\ve q=0)] =\int_0^\infty \D \varepsilon \, g(\varepsilon),
\end{equation}
with
\begin{multline}
g(\varepsilon) = \frac1N \sum_{\boldsymbol\kappa} \delta(\varepsilon-|w(\boldsymbol\kappa)|)\\\times\left[\frac{1}{|w(\boldsymbol\kappa)| + |w(\boldsymbol\kappa+\ve q)|}-\frac{1}{2|w(\boldsymbol\kappa)|} \right]\label{gOfE_definition},
\end{multline}
the integral is convergent for the tight-binding formulation as well as for the Dirac approximation. Furthermore, $g(\varepsilon)$ is only large at low energies, i.e., where the Dirac approximation is valid. Also, the comparison of the Dirac approximation of $g(\varepsilon)$ and the tight-binding calculation in the inset of Fig. \ref{fig_kubo_integrands} shows that the integrand $g(\varepsilon)$ is equal in both calculations.

\subsection{Exact charge response}
In order to scrutinize the results of the linear response calculation we diagonalize the Hamiltonian $H_0 + H_E$ numerically and calculate the charge distribution induced by $H_E$ [Eq. (\ref{ham_lin_response_ext_field})]. Unlike the linear response method, this calculation is not restricted to small fields $E$. We start with considering only the $\pi$ bands in order to be able to check the results of the linear response calculation directly. In a second step, we then take into account all carbon orbitals of the second shell in order to see how the charge response is affected by the $\sigma$ orbitals.

\subsubsection{\texorpdfstring{$\pi$}{pi} band only}
Considering only the $\pi$ band and neglecting all curvature effects, we set $H_0 = H_{\rm hop}^\pi$. We consider only armchair nanotubes and transform the direction along the nanotube to $k$-space. For a nanotube with $N_c$ unit cells in circumferential direction, the Hamiltonian is a $k$-dependent $2N_c\times 2N_c$ matrix, which we diagonalize numerically. From the eigenvalues $\varepsilon_{m,k}$ and the corresponding eigenvectors $\psi_{m,k}(\varphi)$ we calculate the induced charge density (in units of the electron charge $e$)
\begin{equation}
\rho(\varphi) = \frac{2}{N_z}\sum_{m,k} \Theta(\varepsilon_F - \varepsilon_{m,k})|\psi_{m,k}(\varphi)|^2 - 1.\label{numerical_density_pi}
\end{equation}
Note that $\varphi$ must be considered as a discrete variable with $2N_c$ possible values between $0$ and $2\pi$ corresponding to the $A$ and $B$ sublattice sites in the $N_c$ unit cells in circumferential direction. $N_z$ is the number of unit cells along the tube ($z$ direction). The spin-degeneracy is taken into account in Eq. (\ref{numerical_density_pi}).

\begin{figure}[!ht]
\centering
\includegraphics[width=200pt]{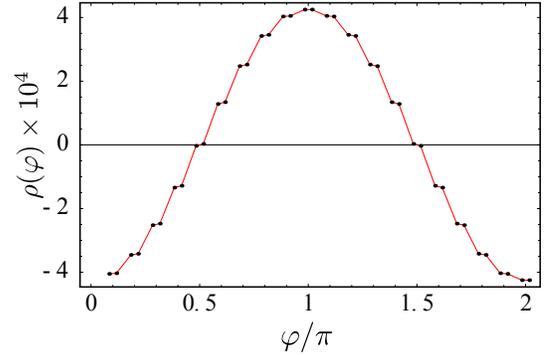}
\caption{(Color online) The induced $\pi$ band charge density $\rho(\varphi)$ for in a (10,10)-CNT for $\phi_{0,\rm tot}=0.01 V_{pp}^\pi$. The black dots show the results of the numerical diagonalization of the $\pi$ band Hamiltonian. The line (red) is the charge response evaluated in linear response. The linear response data is also discrete. It is joined that it can be distinguished from the numerical results. Actually both calculations give the same result.}
\label{fig_pi_band_num_response}
\end{figure}

Fig. \ref{fig_pi_band_num_response} compares the density response to an external field corresponding to $\phi_{0,\rm tot}=0.01 V_{pp}^\pi$, calculated from linear response and from the direct solution of the lattice Hamiltonian. Within numerical accuracy, both calculations give the same results for these small potential amplitudes. Also, one can clearly see the anomalous response, which makes the response curve different from a pure cosine shape.

\begin{figure}[!ht]
\centering
\includegraphics[width=\columnwidth]{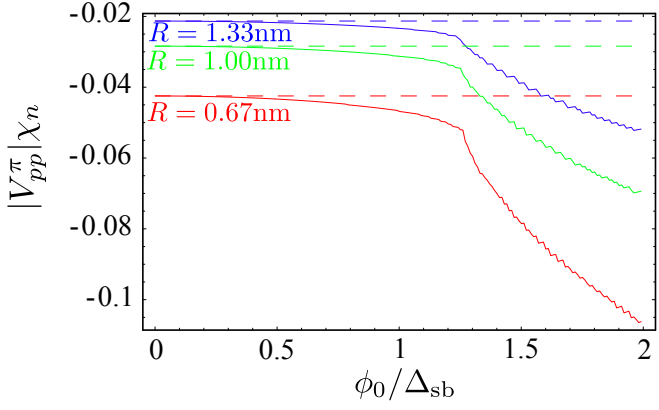}
\caption{(Color online) Nonlinearities in the normal response coefficient $\chi_n$ in the $\pi$ band. The different curves correspond to different nanotube radii $R=0.67,1.00$, and 1.33 nm. The solid lines are the response coefficients fitted to the numerical charge density calculated by Eq. (\ref{numerical_density_pi}). The dashed lines indicate the corresponding linear response coefficients. The applied potential (abscissa) is rescaled with the subband splittings of the nanotubes with the different radii. These are $\Delta_{\rm sb}=(0.31,0.16,0.078)|V^\pi_{pp}|$ for the radii $R=0.67,1.00$, and 1.33 nm, respectively.}
\label{fig_pi_band_nonlin_response}
\end{figure}

The response coefficients $\chi_n$ and $\chi_a$ can be extracted from the numerical calculations of the induced charge densities $\rho(\varphi)$ by a fit to Eq. (\ref{charge_response_form}). The normal response coefficients $\chi_n$ for larger potentials $\phi_{0,\rm tot}$ resulting from these fits are shown in Fig. \ref{fig_pi_band_nonlin_response}. For small external potentials, $\chi_n(\phi_{0,\rm tot})$ is close to its linear response result, Eq. \eqref{ana_normal_response}. For larger potentials, however, there are large deviations. The linear response regime is left if $\phi_{0,\rm tot}$ is on the order of the subband splitting $\Delta_{\rm sb}$. This subband splitting depends on the CNT radius and is given in the caption of Fig. \ref{fig_pi_band_nonlin_response}.

\subsubsection{\texorpdfstring{$\sigma$}{sigma} band contributions}

In the analysis of the Kubo integral we have observed that the contributions to the charge susceptibilities from high energies cannot a priori be neglected. Thus, in order to check the stability of the $\pi$ band calculation of the induced charge density, we calculate the charge redistribution in a tight-binding model containing all second shell orbitals of the carbon atoms, i.e., the $\pi$ and the $\sigma$ bands. For this, we numerically diagonalize the Hamiltonian
\begin{equation}
H = H_{\rm hop} + H_{\rm SO} + H_E,
\end{equation}
with the individual terms of $H$ defined in Sec. \ref{section_model}. The charge density can be separated into a $\pi$ band part and a $\sigma$ band part
\begin{align}
\rho^\pi(\varphi_{\ve n,\zeta}) &= \sum_\lambda \left<c^\dagger_{\ve n \zeta p_r \lambda} c_{\ve n \zeta p_r \lambda} \right> \\
\rho^\sigma(\varphi_{\ve n,\zeta}) &= \sum_\lambda \sum_{\mu=s,p_z,p_t} \left<c^\dagger_{\ve n \zeta \mu\lambda} c_{\ve n \zeta\mu\lambda} \right> .
\end{align}
Both terms have the qualitative form found in the linear response theory. Defining the normal and the anomalous response coefficients for the $\pi$ and $\sigma$ response separately, we find for the exemplary (10,10)-CNT
\begin{align}
|V_{pp}^\pi|\chi_n^\pi &= -0.050 & V_{pp}^\pi\chi_n^\sigma &= -0.0077 \label{1010_chin}\\
V_{pp}^\pi\chi_a^\pi &= 0.011 & V_{pp}^\pi\chi_a^\sigma &= 0.0049\label{1010_chia}
\end{align}
Thus, about 87\% of the normal charge response, which will be important for the field screening discussed in the next subsection, comes from the $\pi$ electrons. Note that the field screening, which enters the Hamiltonian of this numerical calculation, has been taken into account self-consistently.

\subsection{Field screening\label{subsec_screening}}
As mentioned before, the charge response calculated in the previous subsections is not the actual charge response of the carbon nanotube because electron-electron interactions have not been taken into account. Their effect, however, is important because they reduce the amplitude of the charge rearrangement considerably.

In order to see this, we include the electron-electron interactions with a self-consistent Hartree calculation. In addition to this mean-field approximation, we assume that the charge induced by an external electric field is distributed homogeneously on the carbon nanotube surface, i.e., we neglect the anomalous response as it is staggered on the sublattice level and averages out in the continuum approximation. We therefore assume that the electric field induces the continuous charge distribution
\begin{equation}
\rho^{\rm 3D}(\ve r) = \rho_0 \cos(\varphi) \delta(|\ve r|-R), \label{cd_continuum}
\end{equation}
where $R$ is the radius of the nanotube, $\varphi$ is the continuous azimuthal angle of the spatial coordinate $\ve r$ measured from the center of the nanotube, and $\rho_0$ is the amplitude of the induced charge distribution. $\rho_0$ is most easily determined by requiring that the total induced charge in the positive half space ($\cos\varphi >0$) calculated from the continuum approximation Eq. (\ref{cd_continuum}) and from the lattice-resolved expression Eq. (\ref{charge_response_form}) are equal,
\begin{equation}
\int \D^3\ve r \, \Theta(\cos\varphi)\rho^{\rm 3D}(\ve r) = e \sum_{\ve n,\zeta} \Theta(\cos\varphi_{\ve n,\zeta})\rho(\varphi_{\ve n,\zeta}),\label{rho0_condition}
\end{equation}
with $\Theta(x)$ the unit step function and $e$ the electron charge. The anomalous response drops out on the right hand side of Eq. (\ref{rho0_condition}). Furthermore, for the normal response, the right hand side of Eq. (\ref{rho0_condition}) is calculated approximately by using
\begin{equation}
\sum_{\ve n,\zeta} f(\varphi_{\ve n,\zeta}) \simeq \frac La \frac1{\Delta\varphi} \int \D\varphi f(\varphi),
\end{equation}
where $\Delta\varphi = \frac{\sqrt3 a}{4R}$ is the mean azimuthal angle between neighboring lattice sites in the armchair CNT, $L$ is the length of the tube (in $z$ direction) and $f(\varphi)$ is any function. The consequences of the errors, introduced by the above approximation, for the screening properties are much smaller than the consequences of the uncertainties of the hopping parameters. Therefore, we have
\begin{equation}
\rho_0 = \frac{4e}{\sqrt3 a^2} \chi_n \phi_{0,\rm tot},
\end{equation}
with $\phi_{0,\rm tot}$ the amplitude of the electrostatic potential at the tube surface,
including the contributions from the external and the induced fields. This back action effect is the basis for charge screening and will be calculated in the following.

A charge distribution of the form (\ref{cd_continuum}) induces an electric potential $\phi_{\rm ind}$ which can be calculated by
\begin{equation}
\phi_{\rm ind}(\ve r) = \frac{e}{4\pi\varepsilon_0} \int \D^3\ve r' \frac{\rho^{\rm 3D}(\ve r')}{|\ve r - \ve r'|},
\end{equation}
where $\varepsilon_0$ is the vacuum dielectric constant. For symmetry reasons, $\phi_{\rm ind}$ does not depend on the $z$ coordinate (along the tube) but has only a radial ($r$) and azimuthal ($\varphi$) dependence. We find
\begin{equation}
\phi_{\rm ind}(r,\varphi) = \phi_{0,\rm ind} \cos(\varphi ) f(r/R).
\end{equation}
with $\phi_{0,\rm ind} = -\Gamma \phi_{0,\rm tot}$ and
\begin{equation}
f(r/R) = \left\{ \begin{matrix} \frac rR &\text{ for } r<R \\ \frac Rr&\text{ for }r>R\end{matrix}\right..
\end{equation}
The proportionality constant $\Gamma$ is given by
\begin{equation}
\Gamma = 356.78 R[{\rm nm}] |\chi_n| {\rm eV}
\end{equation}
The total electrostatic potential $\phi_{\rm tot}(\ve r)$ felt by the electrons on the tube surface consists of two parts
\begin{equation}
\phi_{\rm tot}(\ve r)=\phi_{\rm ext}(\ve r) + \phi_{\rm ind}(\ve r),\label{total_potential}
\end{equation}
where $\phi_{\rm ext}(\ve r)$ comes from the external electric field and $\phi_{\rm ind}(\ve r)$ is induced by the rearrangement of electron charges at the tube surface.

At the tube surface $|\ve r| = R$, all terms in Eq. (\ref{total_potential}) have the same functional form ($\sim\cos\varphi$), so that the problem of solving the self-consistency equation reduces to an equation for the amplitudes of the electric potentials at the tube surface
\begin{equation}
\phi_{0,\rm tot} = \phi_{0,{\rm ext}} + \phi_{0,\rm ind} = \phi_{0,{\rm ext}} - \Gamma \phi_{0,\rm tot}
\end{equation}
and so
\begin{equation}
\phi_{0,\rm tot} = \phi_{0,{\rm ext}}/\gamma,\bs\bs \gamma= 1+\Gamma.
\end{equation}
Inside the tube ($r<R$), the functional form of $\phi_{\rm tot}(\ve r)$ is the same as $\phi_{{\rm ext}}(\ve r)$, so that the total electric field inside the tube is homogeneous, just as the external field, but is screened by the factor $\gamma$. Thus, we may define the screened field
\begin{equation}
E^* = E/\gamma.
\end{equation}
Note, however, that only inside the nanotube, the field is screened homogeneously. Outside the tube, the total electric field is highly inhomogeneous.

The screening factor $\gamma$ quantifies how free the surface charges are to move. For a metal cylinder $\gamma\rightarrow\infty$, which means that the charges can move freely on the surface. For a cylinder made from an insulating material, the charges are localized and can only form dipoles, which is reflected by $\gamma \gtrsim 1$. A carbon nanotube lies in between the metallic and the insulating limit. For a (10,10)-CNT, for instance, we find $\gamma\simeq 5.6$, where the $\pi$ and $\sigma$ bands are taken into account. Note that $\gamma-1$ depends linearly on the inverse hopping parameters [see, e.g., Eqs. (\ref{1010_chin}) and (\ref{1010_chia})], so that the typical error of $\gamma$ in a tight-binding calculation as performed here can be estimated by the spread of tight-binding parameters found in the literature. Within these error bars of 30\%, our result is in agreement with previous works.\cite{benedict_screening_1995,kozinsky_dft_screening_2006}

\section{Analysis of the low-energy theory\label{sect_analysis_eff_theory}}

\subsection{Spectrum without electric field}
In the absence of electric fields, the CNT is invariant under rotations around its axis and the $z$ projection of the spin $s^z = \pm1$ is a good quantum number. In addition, the valley index $\tau=\pm 1$ is always a good quantum number, as long as there is no intervalley scattering. The term $\hbar v_F\Delta k^z_{\rm cv}\sigma_2$ in $H^{\rm cv}_{\rm orb}$ shifts the momentum of the Dirac point along the tube. Assuming infinitely long tubes, this shift is irrelevant and can be dropped. Thus, we are left with the Hamiltonian
\begin{equation}
H = \tau s^z \beta + \tau \hbar v_F k \sigma_2 + (\hbar v_F (k_t + \Delta k^t_{\rm cv}) + \alpha s^z)\sigma_1,
\label{ham_without_electric_field}
\end{equation}
which is readily diagonalized (see also Ref. \onlinecite{izumida_soi_cnt_2009}). The eigenvalue spectrum has eight branches, given by
\begin{align}
\varepsilon_{u,s_z}(k)&= \tau s^z \beta + \sqrt{(\hbar v_F k)^2 + (\hbar v_F (k^t+\Delta k^t_{\rm cv}) +\alpha s^z)^2},\\
\varepsilon_{d,s_z}(k)&= \tau s^z \beta - \sqrt{(\hbar v_F k)^2 + (\hbar v_F (k^t+\Delta k^t_{\rm cv}) +\alpha s^z)^2},
\label{spectrum_without_electric_field}
\end{align}
where $\tau=\pm1$ and $s^z=\pm1$. In the basis $\left\{\left|A\uparrow\right>,\left|B\uparrow\right>,\left|A\downarrow\right>,\left|B\downarrow\right>\right\}$, the eigenvectors are given by
\begin{equation}
\frac{1}{\sqrt{2}} 
\begin{pmatrix}  1 \\  e^{i \vartheta_+ } \\ 0 \\ 0
\end{pmatrix}, \frac{1}{\sqrt{2}}  
\begin{pmatrix}  -1 \\  e^{i \vartheta_+ } \\ 0 \\ 0
\end{pmatrix},
\frac{1}{\sqrt{2}} \begin{pmatrix}  0 \\  0 \\ 1 \\ e^{i \vartheta_- }
\end{pmatrix},
\frac{1}{\sqrt{2}} \begin{pmatrix}  0 \\  0 \\ -1 \\ e^{i \vartheta_-}
\end{pmatrix},\label{free_eigenvectors}
\end{equation}
with
\begin{equation}
e^{i \vartheta_\pm} = \frac{(\hbar v_F (k^t+\Delta k^t_{\rm cv}) \pm \alpha )+ i \tau \hbar v_F k }{\sqrt{(\hbar v_F (k^t+\Delta k^t_{\rm cv}) \pm \alpha)^2+(\hbar v_F k)^2}}.
\label{vartheta}
\end{equation}

In the case of a semiconductor CNT a gap between the two nearest spin-up and spin-down states is
\begin{equation}
\Delta = 2 |\alpha \pm \beta |,
\end{equation}
where $\pm$ corresponds to electrons and holes, respectively. In the case of an armchair nanotube, the eigenstates are two-fold spin-degenerate and the gap between holes and electrons is $ 2 |\alpha |$. 

An electric field perpendicular to the tube axis gives rise to the spin-orbit term $H_{\rm SO} \propto S^y$, and so $s^z$ no longer is a good quantum number. In the following we discuss the consequences of such an electric field.

\subsection{Helical modes}

As already discussed in Ref. \onlinecite{klinovaja_helical_modes_2011}, the interplay of strong electric fields and spin-orbit interaction leads to helical modes. For an armchair CNT the chiral angle is $\theta=\pi/6$ and all terms proportional to $\cos3\theta$ vanish, i.e., $\Delta k^t_{\rm cv}=0$ and $\beta=0$. Furthermore the longitudinal $k$-space shift $\Delta k_{\rm cv}^z$ can be ignored since it can be removed by regauging the phase of the orbitals. This leads to the effective Hamiltonian for the lowest subband in an armchair CNT
\begin{equation}
H^{\rm arm}=\tau \hbar \upsilon_F k \sigma_2 + \alpha S^z\sigma_1 + \tau e E \xi S^y \sigma_2,
\label{ham_armchair}
\end{equation}
which has four branches of eigenvalues
\begin{equation}
\varepsilon(k) = \pm e E \xi \pm \sqrt{\alpha^2 + (\hbar \upsilon_F k )^2}
\label{eq:spectrum_armchair}
\end{equation}
for each valley. In the basis $\left\{\left|A\uparrow\right>,\left|B\uparrow\right>,\left|A\downarrow\right>,\left|B\downarrow\right>\right\}$, the corresponding eigenvectors are given by
\begin{equation}
\frac{1}{2} 
\begin{pmatrix}  -1 \\ e^{i \varsigma } \\ e^{i \varsigma } \\ 1
\end{pmatrix}, \frac{1}{2} 
\begin{pmatrix}  1 \\  -e^{i \varsigma } \\ e^{i \varsigma } \\ 1
\end{pmatrix},
\frac{1}{2}\begin{pmatrix}  1 \\ e^{i \varsigma } \\ -e^{i \varsigma } \\ 1
\end{pmatrix},
\frac{1}{2}\begin{pmatrix}  1 \\  e^{i \varsigma } \\ e^{i \varsigma } \\ -1
\end{pmatrix},\label{helical_eigenvectors}
\end{equation}
with
\begin{equation}
e^{i \varsigma } =  \frac{\alpha + i \tau\hbar \upsilon_F  k}{\sqrt{\alpha^2+(\tau\hbar \upsilon_F  k)^2}}.
\end{equation}
Note that the eigenvectors are independent of the electric field $E$; only the energetical order depends on $E$. In the following, we label the four branches for each valley by $n=1,...,4$. For each $k$, $n=1$ corresponds to the highest eigenvalue and $n=4$ to the lowest.

\begin{figure}[!ht]
\centering
\includegraphics[width=190pt]{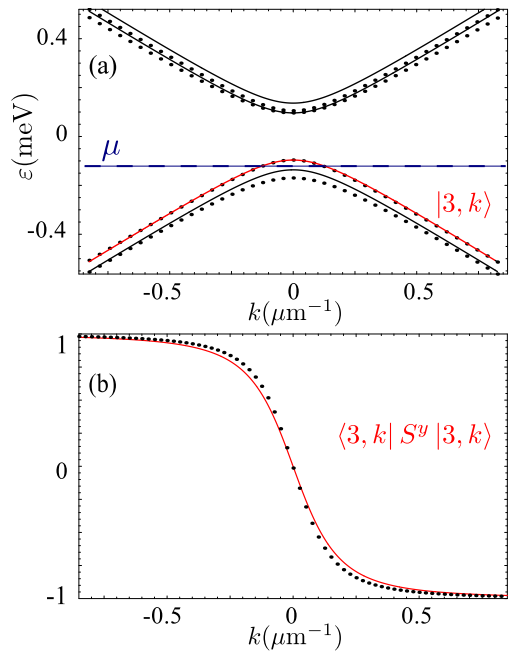}
\caption{(Color online) (a) Low-energy spectrum for a (10,10)-CNT (armchair) in a field $E=1$ V/nm. (b) The dependence of $y$-spin polarization $\left<3, k\right|S^y\left|3, k\right>$ on the momentum $k$ along the tube. The solid lines are the results of the analytical low-energy effective theory. The dots correspond to numerical calculations (see text).}
\label{fig_helical_modes_10}
\end{figure}

\begin{figure}[!ht]
\centering
\includegraphics[width=190pt]{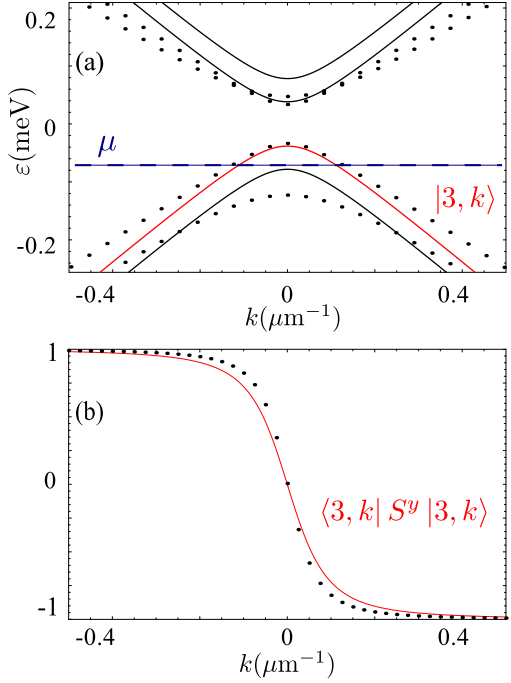}
\caption{(Color online) a) Low-energy spectrum for armchair CNT (20,20) in a field $E=1$ V/nm. b) The dependence of $y$-spin polarization $\left<3,k\right|S^y\left|3,k\right>$ on a wave vector $k$. The solid lines are the results of the analytical low-energy effective theory. The dots correspond to numerical calculations (see text).}
\label{fig_helical_modes_20}
\end{figure}

\begin{figure}[!ht]
\centering
\includegraphics[width=200pt]{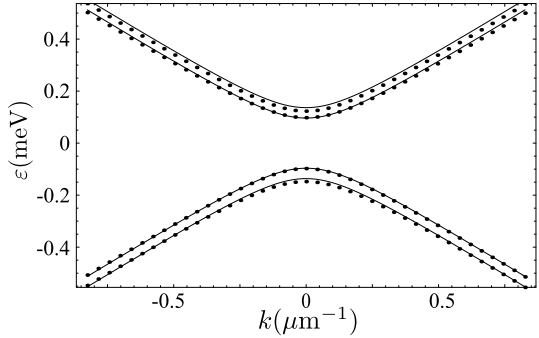}
\caption{(Color online) Low-energy spectrum for armchair CNT (10,10) in a field $E=1$ V/nm assuming that the screening factor $\gamma$ is equal to 20. The solid lines are the results of the analytical low-energy effective theory. The dots correspond to numerical calculations. In this limit we observe a good agreement between two models, which confirms our hypothesis that discrepancies in Fig. \ref{fig_helical_modes_10} are caused by the fact that we neglected the change in charge distribution caused by electric field.}
\label{fig_screening_20}
\end{figure}

In Figs. \ref{fig_helical_modes_10}(a) and \ref{fig_helical_modes_20}(a), the spectrum calculated from the analytical low energy theory is compared to the spectrum calculated from the numerical solution of the complete tight-binding model [Eq. (\ref{full_hamiltonian})]. Obviously, there are significant differences. These are due to the neglect of the Hamiltonian $H_E^{(1)}$ in the derivation of the analytical theory. It is known\cite{novikov_vF_renormalization_2006} that $H_E^{(1)}$ alone gives rise to a renormalization of the Fermi velocity on the order of the squared screened electric field $(E^*)^2=(E/\gamma)^2$. However, the combination of $H_E^{(1)}$ with other parts in the spin-orbit Hamiltonian which depend trigonometrically on $\varphi$ may give rise to effects of first order in $E^*\sim 1/\gamma$. Thus, the differences of the numerical and analytical spectrum should disappear in the limit of perfect screening, i.e. $\gamma\rightarrow\infty$. In order to check this expectation, we increase the screening parameter $\gamma$ in the numerical calculation and find that the agreement between numerics and analytics becomes better for larger screening. For $\gamma=20$, the analytical theory is as good as the numerics, as shown in Fig. \ref{fig_screening_20}. Better screening can be achieved, for instance, by filling the CNT with a dielectric.

The reason why these effects are important although they are of higher than second order is that the parameter regime we are discussing here is at the border of the applicability of perturbation theory. Nevertheless, as we will argue now, the effective theory is valuable also in this regime because it correctly captures the most important features of the helical modes. Moreover, we will show that the numerical results predict an even better spin-polarization than the analytical theory.

As mentioned above, the actual value of the parameter $\xi$ is unknown. We use here a conservative estimate $\xi=2\cdot 10^{-5}$, but exhausting the range of values for $\xi_0$, $\Delta_{\rm SO}$, etc., $\xi$ can increase by one order of magnitude.\footnote{Note that in Table I of Ref. \onlinecite{klinovaja_helical_modes_2011}, $\xi$ was given an order of magnitude too big by mistake. However, given the uncertainty of the parameters $\xi_0$, $\Delta_{\rm SO}$, etc., and thus of $\xi$, all results remain valid.} In this case, i.e. for $\xi=2\cdot 10^{-4}$, the analytical theory fits much better to the numerical calculations. This is essentially because the spin-splitting generated by $eE\xi S^z \sigma_2$ is sufficiently large so that the higher order terms, causing the deviations from the simple analytical model, are not effective in this case. We provide a movie in the EPAPS which illustrates this.\cite{EPAPS} In this movie, the numerical spectrum is compared to the analytical spectrum as the parameter $\xi_0$ is varied between 0.5\AA\, and 0.05\AA. It is seen that the agreement is better for larger $\xi_0$.

Next, we discuss the spin-polarization. The eigenstates given in Eq. (\ref{helical_eigenvectors}) allow us to calculate the spin-polarization of each branch in the analytical model. It is easily seen that $\left<k,n\right|S^{x,z}\left|k,n\right>=0$, i.e., the spin is only polarized in the $y$ direction (perpendicular to the CNT axis and to the electric field). In this sense, the spin is perfectly polarized, even though $\left<S^y\right>$ is smaller than one; the vector $\left<\ve S\right>$, with $\ve S=(S^x,S^y,S^z)$, is perfectly parallel to the $y$ direction and has no components perpendicular to $\hat{\ve y}$. This $y$-spin polarization in the analytical model is given by
\begin{equation}
\left<k,n\right|S^y\left|k,n\right>= \pm \frac{k}{\sqrt{k^2+(\alpha/\hbar v_F)^2}} .
\label{eq:polarization}
\end{equation}
Note that $\left<k,n\right|S^y\left|k,n\right>$ is odd in $k$. This means that for Fermi levels as indicated in Figs. \ref{fig_helical_modes_10} and \ref{fig_helical_modes_20}, we have one helical liquid per Dirac point. The sign of the helicity, however, is the same for each Dirac point (this is a consequence of time-reversal invariance), so that a CNT with a properly tuned Fermi level is a perfect spin filter.

In Figs. \ref{fig_helical_modes_10}(b) and \ref{fig_helical_modes_20}(b), the analytical spin-polarizations are compared with the numerical spin-polarizations. They agree well and the analytical spin-polarization is seen to be a lower bound to the numerical result. One should also note that, because of the increased splitting at zero $k$ between states 3 and 4 of the numerical calculation compared to the analytical results, the range of possible Fermi levels for the helical liquid is increased. This leads to a higher maximum $\left<S^y\right>$ in the more rigorous numerical solution. Also in this sense, the analytical model provides a lower bound on the maximum $\left<S^y\right>$. For example, in the case of a (10,10)-CNT to which an electric field $E=1$V/nm is applied (see Fig. \ref{fig_helical_modes_10}) one can achieve $\left<S^y\right>\simeq 90\%$ in the helical phase.

Another well studied effect of the Hamiltonian $H_E^{(1)}$ is a renormalization of the Fermi velocity. We have not taken this renormalization into account in the analytical calculation in Fig. \ref{fig_helical_modes_20}, but it is accounted for in the numerical calculation. Also this effect allows us to go to larger $k_F$ in the helical regime.

Now that we have understood which features are well captured by the analytical model (e.g., the spin-polarization and the spectrum of the holes), and which are not described correctly (the electron spectrum for large fields), we discuss the case of non-armchair nanotubes on the basis of the analytical model. As mentioned above, the effects leading to deviations from the analytical results can be suppressed by increasing the screening of the externally applied electric field.

For non-armchair, but metallic nanotubes (i.e., $k_t=0$), the chiral angle is $\theta\neq \pi/6$ and $\cos(3\theta)\neq0$. This gives rise to two additional terms in the Hamiltonian [see Eq. (\ref{ham_armchair})]. One term, $\hbar v_F \Delta k_{\rm cv}^t\sigma_1$, opens an orbital gap. The other term, $\tau\beta S^z$, acts as an effective Zeeman field along the tube axis. For the mechanism of valley suppression, discussed in the next subsection, it is important to note that both additional terms have opposite signs in different valleys $\ve K,\ve K'$.

\begin{figure}[!ht]
\centering
\includegraphics[width=200pt]{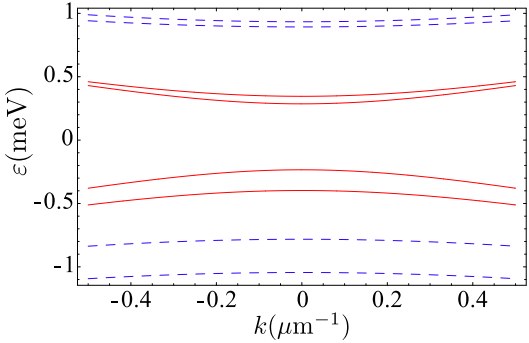}
\caption{(Color online) Low-energy spectrum for two chiral metallic CNTs with different chiralities but comparable radii. The field strength is 0.5 V/nm. The four bands with smaller absolute energy (red, solid) correspond to a (23,20)-CNT with $R=1.44$nm and $\theta=0.154\pi$. The four bands with larger absolute energy (blue, dashed) are from a (26,17)-CNT with $R=1.44$nm and $\theta=0.128\pi$.}
\label{fig_spectrum_chiral_nanotube}
\end{figure}

The most prominent effect of $\theta\neq \pi/6$ is the opening of a large orbital gap in the meV range for reasonable CNT radii $R$. For a nanotube with $R=1.44$ nm but different helical angles, the low-energy spectrum is shown in Fig. \ref{fig_spectrum_chiral_nanotube}. In the armchair CNT, the helical liquid appears on an energy scale of a few hundred $\mu$eV. Thus, it is not a priori clear that the helicity of the left/right movers survives the departure from the pure armchair topology of the lattice. However, a plausibility argument for the stability of the helicity with respect to $\hbar v_F \Delta k_{\rm cv}^t\sigma_1$ can be given: for large $k\gg\Delta k_{\rm cv}^t$, the electronic states are eigenstates of the operator $\sigma_2 S^y$. In this limit the $y$-spin polarization (i.e., the helicity) becomes 100\%. The $y$-spin polarization changes sign under $k\rightarrow-k$. Furthermore, as long as there is no band crossing, $\left<S^y\right>$ must be a smooth, odd function of $k$. This means, as long as $|k_F|\neq 0$, the $y$-spin polarizations of the second band at $\pm k_F$ must be opposite. The question is only, how large is the amplitude of the polarization.

\begin{figure}[!ht]
\centering
\includegraphics[width=200pt]{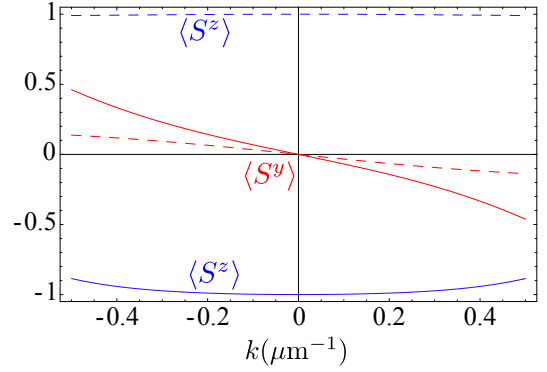}
\caption{(Color online) Spin-polarizations of the third band in $y$ and $z$ direction in chiral nanotubes as functions of $k$. The field strength is 0.5 V/nm. The odd functions around $k=0$ (red) correspond to $\left<S^y\right>$ and the even functions (blue) to $\left<S^z\right>$. The solid lines are results from a (23,20)-CNT and the dashed lines from a (26,17)-CNT.}
\label{fig_spin_polarization_chiral_nanotube}
\end{figure}

The orbital gaps are not the only effect of a nontrivial chirality $\theta\neq\pi/6$. The effective Zeeman field $\tau\beta S^z$ leads to an additional spin-polarization in $z$ direction (along the tube). This can be observed in Fig. \ref{fig_spin_polarization_chiral_nanotube}. At small $k$, where the electronic states are not forced into eigenstates of $\sigma_2$, the effect of this effective Zeeman field is largest, i.e., the $z$ polarization is maximal, while, at large $|k|$, the spin tends to be aligned in $y$ direction. This effect reduces the quality of the spin helicity in that the spin alignment is not completely odd in $k$ and not along $\hat{\ve y}$. Only the $y$-spin component is odd, but the $z$-spin is an even function of $k$.

\subsection{Valley suppression}

In the previous section we have seen how the quality of the helicity is reduced in case of a non-armchair chirality of the CNT. This was due to an orbital $k$-shift $\hbar v_F \Delta k_{\rm cv}^t\sigma_1$ and an effective Zeeman field $\tau\beta S^z$, both of which are consequences of the chiral angle $\theta\neq \pi/6$ deviating from the armchair case. Here, we show that the appearance of these terms can be turned into an advantage: by applying an additional magnetic field along the tube, the perfect helicity can be restored in one valley while all bands of the other valley are removed from the low-energy regime.

First note that both, the orbital $k$-shift and the effective Zeeman field, can be interpreted as originating from a magnetic field applied along the axis of the CNT. In general, a magnetic field $\ve B$ has two effects on the electrons. First, it induces a Zeeman energy, described by the Hamiltonian
\begin{equation}
H_Z = \mu_B \ve B \cdot \ve S,
\end{equation}
where $\mu_B$ is the Bohr magneton and $\ve S$ is the vector of Pauli matrices for the electron spin (eigenvalues $\pm1$). Second, if the magnetic field has a component along the CNT axis, i.e., $B_z\neq 0$, the transverse wave function of the electron encloses magnetic flux and this gives rise to a shift in the electron momentum in circumferential direction
\begin{equation} \label{shift_by_Bz}
\Delta k^t_{B} = \frac{\pi B_z R}{\Phi_0},
\end{equation}
where $\Phi_0=h/|e|$ is the magnetic flux quantum. Since $\Delta k^t_{B}$ must be added to $\Delta k^t_{\rm cv}$, as well as $\mu B_z S^z$ must be added to $\tau\beta S^z$, the effect of a non-armchair chirality of the CNT can be compensated for by a magnetic field, at least to a certain degree as explained below. However, since the chirality-induced $\Delta k^t_{\rm cv}$ and $\tau\beta$ have opposite signs in different valleys, but the real magnetic field terms $\Delta k_B^t$ and $\mu_B B_z$ have not, this compensation works only in one of the two valleys. In the other valley, instead, the effect of the chirality is even increased.

Note that in general only one of the two chirality effects can be compensated for by a magnetic field $B_z$. Compensating for the orbital gap requires
\begin{equation}
\hbar v_F\frac{\pi B_z R}{\Phi_0} = \tau \frac{5.4{\rm meV}}{R[{\rm nm}]^2}\cos3\theta,\label{compensate_orbital_gap}
\end{equation}
while compensating for the effective Zeeman field requires
\begin{equation}
\mu_B B_z = \tau \frac{0.31{\rm meV}}{R[{\rm nm}]}\cos3\theta.\label{compensate_eff_zeeman}
\end{equation}
In general, Eqs. (\ref{compensate_orbital_gap}) and (\ref{compensate_eff_zeeman}) are not compatible. 

\begin{figure}[!ht]
\centering
\includegraphics[width=220pt]{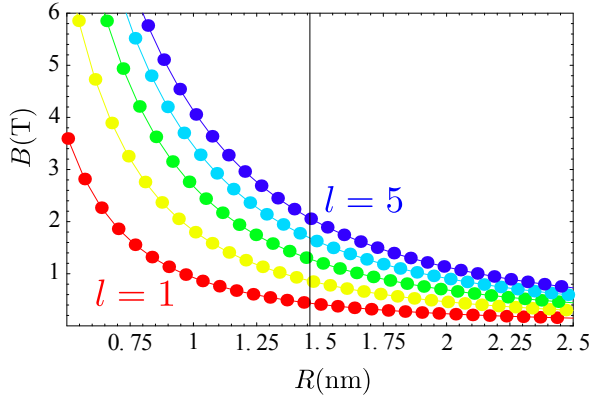}
\caption{(Color online) Magnetic fields needed for compensating the chirality-induced effective Zeeman term in different $(m+3l,m)$-CNTs as a function of CNT radius $R$. $l$ varies from 1 (lower curves, red) to 5 (upper curves, blue). The dots correspond to the discrete radii and chiralities. The interconnecting lines are guides to the eye. The vertical line shows the optimal CNT radius at which both, the Zeeman term and the orbital shift, are compensated simultaneously.}
\label{fig_chiralities_and_compensating_magnetic_fields}
\end{figure}

However, these two conditions have a different dependence on the CNT radius $R$, so that there exists an optimal radius $R_{\rm opt}\simeq 1.46$ nm at which  (\ref{compensate_orbital_gap}) and (\ref{compensate_eff_zeeman}) are compatible. For $R=R_{\rm opt}$, the compensating magnetic field is
\begin{equation}
B_{z,\rm opt} = 3.67 T \cdot \cos3\theta.
\end{equation}
Of course, the CNT radius is not a continuous variable, but can only take on discrete values. Fig. \ref{fig_chiralities_and_compensating_magnetic_fields} shows the fields needed for compensating the effective Zeeman term for different CNT chiralities. We see in Fig. \ref{fig_chiralities_and_compensating_magnetic_fields} that there are indeed CNTs that have an optimal radius. In fact, the uncertainty in determining the optimal radius of actual CNTs is larger than the spacing of the dots in Fig. 14 so that all dots in the vicinity of the vertical line in the figure can be considered as optimal. 

\begin{figure}[!ht]
\centering
\includegraphics[width=245pt]{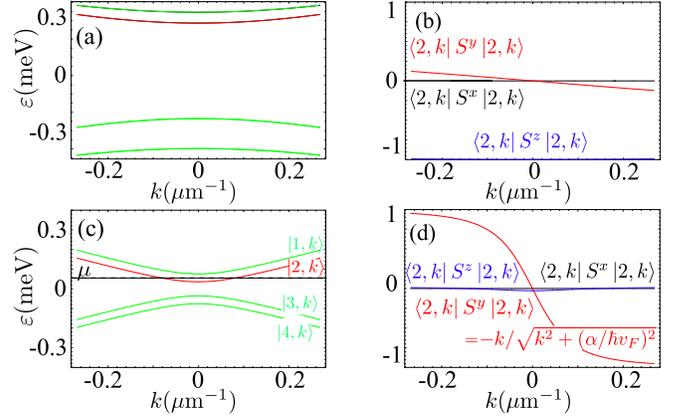}
\caption{(Color online) Chiral (23,20)-CNT with electric field $E=1$V/nm. 
(a) The spectrum and (b) the spin expectation values at the $\ve K/ \ve K'$ point for the $\left|2,k\right>$ subband and magnetic field $B_z=0$.
For $B_z=0.46$T, the bands (c) at $\ve K'$ (not shown) are gapped, while the spectrum at $\ve K$ (solid lines) has the same form as in the armchair case [Eq. (\ref{eq:spectrum_armchair})]. The size of the gap is $2 |\alpha|=0.16$meV. 
 The spin expectation values (d) at $\ve K$ for $\left|2,k\right>$ follow closely the armchair case [see Eq. (\ref{eq:polarization})].}
\label{fig_valley_suppression}
\end{figure}

\subsection{External magnetic fields in armchair CNTs}

In this section we discuss the low-energy spectra of armchair carbon nanotubes in strong electric fields and additional magnetic fields $\ve B$ along the three possible spatial directions.

\subsubsection{Magnetic field along the nanotube}

\begin{figure}[!ht]
\centering
\includegraphics[width=\linewidth]{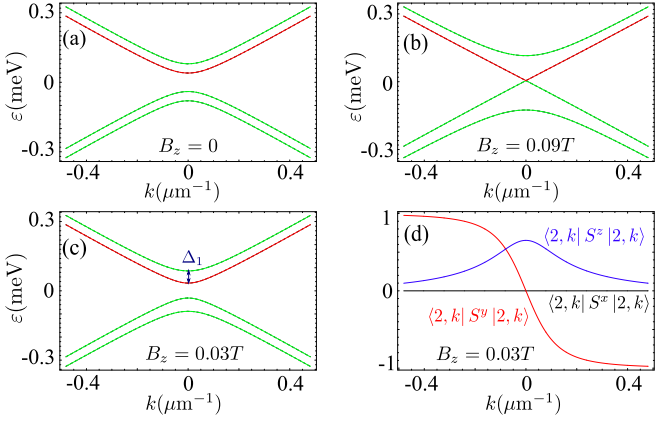}
\caption{(Color online) Spectrum of a (20,20)-CNT in an electric field $E= 1 V/nm$ and different magnetic fields $B_z=0, 0.09, 0.03$ T (parts a,b,c) along the nanotube axis ($z$ direction). The dependence of the spin-polarization on the wave vector $k$ is shown in part d) for $B_z=0.03$ T . In the limit of the large magnetic fields the Zeeman term dominates and  $\left<S^z\right>\rightarrow \pm 1$. See Eq. (\ref{k0_splitting}) for $\Delta_1$.}
\label{fig_Magnetic_Field_Along_Nanotube}
\end{figure}

A magnetic field along the CNT axis $\ve B = B_z \ve{\hat z}$ leads to the shift of the circumferential wave vector, given by Eq. \eqref{shift_by_Bz}. Together with the Zeeman term, this leads to the additional term in the Hamiltonian
\begin{equation}
H_{\rm mag} = \hbar \upsilon_F \frac{ \pi B_z R}{\Phi_0} \sigma_1 + \mu_B B_z S^z.
\label{magnetic_along}
\end{equation}
The resulting spectrum of a (20,20)-CNT in a 1 V/nm electric field is shown in Fig. \ref{fig_Magnetic_Field_Along_Nanotube} for different magnetic field strengths. 

As explained above, for the case of a magnetic field along the CNT axis, the additional terms in the Hamiltonian can be accounted for by redefining the parameters $\beta$ and $\Delta k^t_{\rm cv}$, i.e.,
\begin{align}
\beta^* &= \beta+ \mu_B B_z, \label{eq:beta} \\
\Delta k^{t*} &= \Delta k_{\rm cv}^{t}+\pi B_z R/ \Phi_0. \label{eq:k_cv}
\end{align}
The energy spectrum at $k=0$ for a chiral nanotube is given by
\begin{align}
\varepsilon_{1,3}&= - \alpha \pm \sqrt{(eE\xi)^2+(\beta^*-\hbar \upsilon_F \Delta k^{t*})^2}, \\
\varepsilon_{2,4}&= \alpha \pm \sqrt{(eE\xi)^2+(\beta^*+\hbar \upsilon_F \Delta k^{t*})^2}.
\end{align}
The splitting $\Delta_1$ at $k=0$ is given, to leading order in the magnetic field $B_z$ and for an armchair nanotube, by
\begin{align}
\Delta_1 &\simeq 2 \left |e E \xi  +  \frac{\left(\beta^*+\hbar \upsilon_F \Delta k^{t*}\right)^2 }{ 2 e E \xi  } \right|.\label{k0_splitting}
\end{align}

\subsubsection{Magnetic field along the electric field}
\begin{figure}[!ht]
\centering
\includegraphics[width=\linewidth]{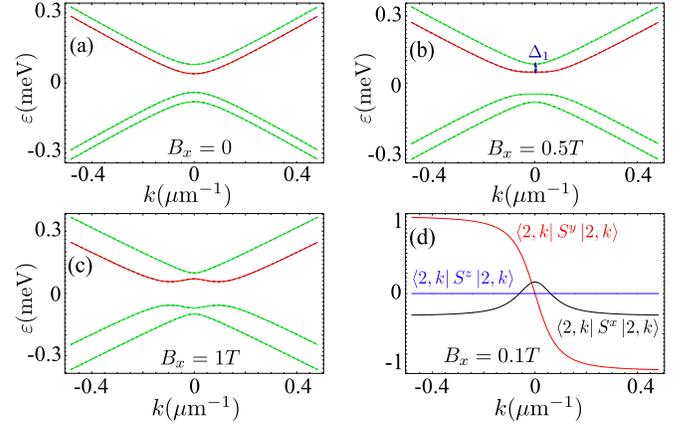}
\caption{(Color online) The spectrum of a (20,20)-CNT in the electric field $E= 1 $V/nm and different magnetic fields $B_x=0, 0.5 , 1$ T (parts a,b,c) along the electric field ($x$ direction). The dependence of the spin-polarization on the wave vector $k$ is represented on the graph d) for $B_x=0.1$ T. In the limit of large magnetic fields the Zeeman term dominates and  $\left<S^x\right>\rightarrow \pm 1$. See Eq. (\ref{k0_splitting2}) for $\Delta_1$.}
\label{fig_Magnetic_Field_Along_Electric_Field}
\end{figure}

In case of a magnetic field perpendicular to the CNT axis we take into account only the Zeeman term, as the orbital effect is small for the strengths of the fields considered here. For $\ve B=B_x \hat{\ve x}$ parallel to the electric field the additional term in the Hamiltonian reads
\begin{equation}
 H_{\rm mag} = \mu_B B_x S^x.
\end{equation}
For an armchair CNT, the total Hamiltonian can then be diagonalized analytically. We find
\begin{multline}
\varepsilon = \pm \biggl[(eE\xi )^2+\alpha ^2+(\mu_B B_x)^2+ (\hbar \upsilon _F k)^2\\\pm2 \sqrt{(eE\xi \alpha )^2  + \left((eE\xi )^2+(\mu_B B_x) ^2\right) (\hbar \upsilon _F k)^2}\biggr]^{\frac12}.
\end{multline}
In Fig. \ref{fig_Magnetic_Field_Along_Electric_Field} the spectrum is shown for several values of the magnetic field. $B_x$ decreases the splitting at $k=0$
\begin{eqnarray}
\Delta_1 &\simeq& 2 \left | e E \xi \left(1-\frac{1}{2}\frac{\mu_B^2B_x^2}{\alpha^2-(e E \xi)^2}\right) \right |.\label{k0_splitting2}
\end{eqnarray}

\begin{figure}[!ht]
\centering
\includegraphics[width=\linewidth]{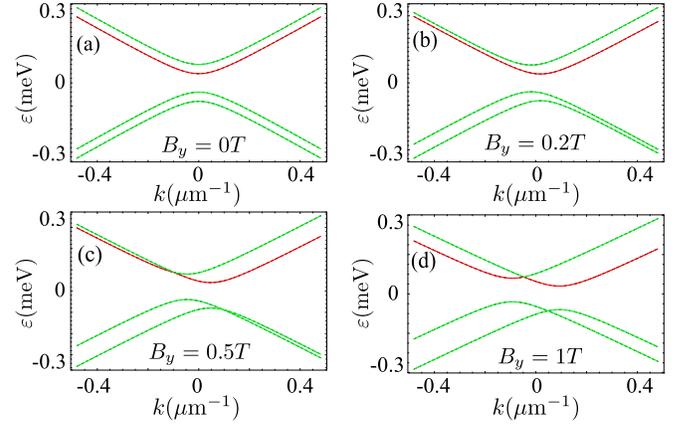}
\caption{(Color online) The spectrum of the CNT (20,20) in the electric field $E= 1 V/nm$ and different magnetic fields $B_y=0, 0.2, 0.5, 1 $ T (parts a,b,c,d) perpendicular to the electric field and to the nanotube axis ($y$ direction). The dependence of the spin-polarization on the wave vector is given by Eq. (\ref{eq:polarization}) with the shifted wave vector.}
\label{fig_Magnetic_Field_Perpendicular_Electric_Field}
\end{figure}

\subsubsection{Magnetic field perpendicular to the electric field and the nanotube axis}
A magnetic field perpendicular to the electric field and to the nanotube axis is described by the Zeeman term
\begin{equation}
 H_{mag} = \mu_B B_y S^y.
\end{equation}
In this case the magnetic field tries to polarize the spin in the same direction as the electric field. For an armchair nanotube the Hamiltonian can be diagonalized exactly giving the energy spectrum
\begin{eqnarray}
\varepsilon_{1,2}&=&- e E \xi \pm \sqrt{ \alpha^2+ (\hbar \upsilon_F k-\mu_B B_y)^2},\\
 \varepsilon_{3,4}&=& e E \xi \pm \sqrt{\alpha^2+ (\hbar \upsilon_F k +\mu_B B_y)^2},
\end{eqnarray}
which is shown in Fig. \ref{fig_Magnetic_Field_Perpendicular_Electric_Field}. $B_y$ leads to an additional shift $\Delta k=\pm\mu_B B_y/\hbar \upsilon_F$ of the momentum along the nanotube axis.

\section{Resonant spin transitions\label{section_edsr}}

The manipulation of the electron spin by time-dependent external fields is most important for quantum computing and spintronics. Traditionally, a time-dependent magnetic field, coupling to the electron spin via the Zeeman energy, is utilized for Rabi flopping.\cite{atherton_esr_book} However, the combination of time-dependent electric fields and spin-orbit coupling may give rise to an all-electric control of the electron spin.\cite{kato_edsr_2003,golovach_edsr_2006,nowack_edsr_2007,bulaev_edsr_2007, laird_edsr_2009} This effect is called electric dipole spin resonance (EDSR) and it was argued semiclassically\cite{bulaev_soi_cnt_2008} that this concept is applicable to CNTs in principle. In the following, we investigate the EDSR effect due to $H^{\rm el}_{\rm SO}$, which has been derived microscopically in Sec. \ref{sect_effective_hamiltonian}.

In a spin resonance experiment one considers a spin which is split by a Zeeman energy $E_Z$. In order to drive transitions between the spin-up and spin-down states, an external field with frequency $\omega = E_Z/\hbar$ is required. In the case of a CNT, there are two regimes for such transitions, characterized by the frequencies required to drive them. Spin resonance at optical frequencies (THz regime) involve different subbands. They are possible \footnote{The electromagnetic field, described by the time-dependent vector potential
$\ve A_{ac}( \ve r,t) = \ve A_0 \sin (\omega t - \ve k \cdot \ve r)$, leads to inter-subband transitions $H^{ind-tr}= \upsilon_F e A_0 i(n^+-n^-)/2c$. The interplay of $H^{ind-tr}$ with the spin-orbit interaction $H_{\pi}^\mathrm{SO-tr} = \gamma_1 \sigma_3 ( S^- n^++S^+ n^-)+ \tau 
(\beta_1 \sin 3 \theta+\alpha \sigma_2) i (S^+ n^--S^- n^+)$, where the subband transitions come from the angular dependencies of the spin operators in Eq. (\ref{h_pi_angle_dependent}), enables the realization of electric dipole spin resonance at optical frequencies. This effect, however, is beyond the scope of this work.
} but not of interest in this work. The intra-subband transitions have characteristic energy scales below 1 meV, which corresponds to frequencies in the GHz regime. AC voltages in the GHz regime can easily be generated electronically, so that this regime is suitable for EDSR. 

We now discuss the transitions between electronic states with opposite spins in the lowest subband, induced by a time-dependent electric field
\begin{equation}
\ve E_{ac}(t) = \ve E_{ac} \cos \omega t.
\end{equation}
The effective Hamiltonian describing the interaction with a time-dependent field is similar to Eq. (\ref{ham_electric_field}). However, we assume $\ve E_{ac}$ along the $y$ direction, so that
\begin{equation}
H^{\rm ac}(t) = \tau e E_{ac}(t) \xi S^x \sigma_2.
\label{E_along_OY}
\end{equation}
This term can be used to implement the EDSR effect. The frequency $\omega$ of the field is chosen to fit the energy difference between two eigenstates of $H_{\pi}^{\rm eff}$ [Eq. (\ref{eq:effective_hamiltonian})], between which the transitions are induced.

Indeed, the form of $H^{\rm ac}(t)$ is typical for the EDSR effect. It has a trigonometric time dependence and is proportional to the spin operator $S^x$. However, $H^{\rm ac}(t)$ is also proportional to the sublattice operator $\sigma_2$, and this leads to an additional complication compared to a simple spin resonance Hamiltonian of the form $\cos(\omega t) S^x$. This complication disappears in the limit of large $k$ where the term $\tau\hbar v_F k\sigma_2$ dominates in $H_{\pi}^{\rm eff}$ and one can assume $\sigma_2$ to be a good quantum number, i.e., $\sigma_2=\pm 1$. In this case the Rabi frequency is given by 
\begin{equation}
\omega_{R}^{*} = \frac{eE_{ac}\xi}{\hbar}.
\label{Rabi_limit}
\end{equation} 
The occurence of the sublattice operator in the coupling term reduces the Rabi frequency $\omega_{R}<\omega_R^*$ around $k\simeq 0$.

\subsection{Without dc electric field}
We start with considering non-armchair CNTs with $\theta\neq\pi/6$. In this case, transitions between the states $\varepsilon_{u,+}$ and $\varepsilon_{u,-}$ [see Eq. (\ref{spectrum_without_electric_field})] are driven by $H^{\rm ac}(t)$. For the Rabi frequency we obtain
\begin{equation}
 \omega_{R}=\frac{eE_{ac}\xi}{\hbar}\left |\sin(( \vartheta_++ \vartheta_-)/2)\right|,\label{rabi_frequency_no_dc_field}
\end{equation}
where $\vartheta_{\pm}$ is defined in Eq. (\ref{free_eigenvectors}). In agreement with what was explained above, Eq. (\ref{rabi_frequency_no_dc_field}) reduces to $\omega_R^*$ in the limit $k\rightarrow\pm\infty$.

\begin{figure*}[!hbt]
\begin{center}
\includegraphics[width=\linewidth]{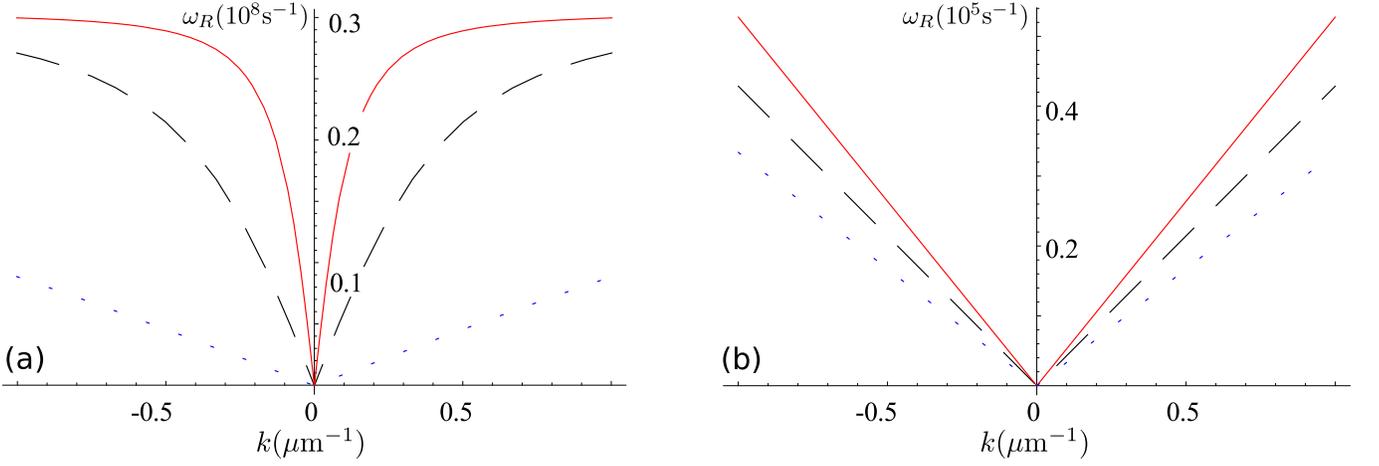}
\caption{(Color online) The dependence of the Rabi frequency $\omega_R$ on the wave vector $k$ along the CNT axis (a) for the metallic nanotubes (13,10)-CNT (dotted), (23,20)-CNT (dashed), and (33,30)-CNT (solid). (b) shows $\omega_R$ for semiconducting nanotubes:  the CNT (12,10) - dotted,  the CNT (18,10) - dashed,  the CNT (24,10) - solid. The amplitude of the ac electric field is 1 mV/nm.}
\label{fig_Rabi_from_wavevector_metal_semiconductor}
\end{center}
\end{figure*}

For $k=0$, the wavefunction is an eigenfunction of $\sigma_1$. The states $\varepsilon_{u,+}$ and $\varepsilon_{u,-}$ have opposite spins but the same isospins $\sigma_1$, if $\hbar \upsilon_F (k^t+ \Delta k^t_{\rm cv}) > \alpha$. The time-dependent electric field [Eq. (\ref{E_along_OY})], however, couples spin ($S$) and isospin ($\sigma$) simultaneously. Thus, at $k=0$ Rabi flopping between $\varepsilon_{u,+}$ and $\varepsilon_{u,-}$ via $H^{\rm ac}(t)$ is not allowed and the Rabi frequency is zero (see Fig. \ref{fig_Rabi_from_wavevector_metal_semiconductor}). Near the Dirac points the Rabi frequency is proportional to $k$. We find
\begin{equation}
\omega_{R}\simeq\frac{eE_{ac}\xi}{\hbar}\left | \frac{k}{k^t+ \Delta k^t_{\rm cv} } \right|.
\label{omega_small_k}
\end{equation}
For non-metallic CNTs, $k_t$ is very large and this leads to a strong suppression of the Rabi frequency (see Fig. \ref{fig_Rabi_from_wavevector_metal_semiconductor}).

\begin{figure}[!hbt]
\centering
\includegraphics[width=\linewidth]{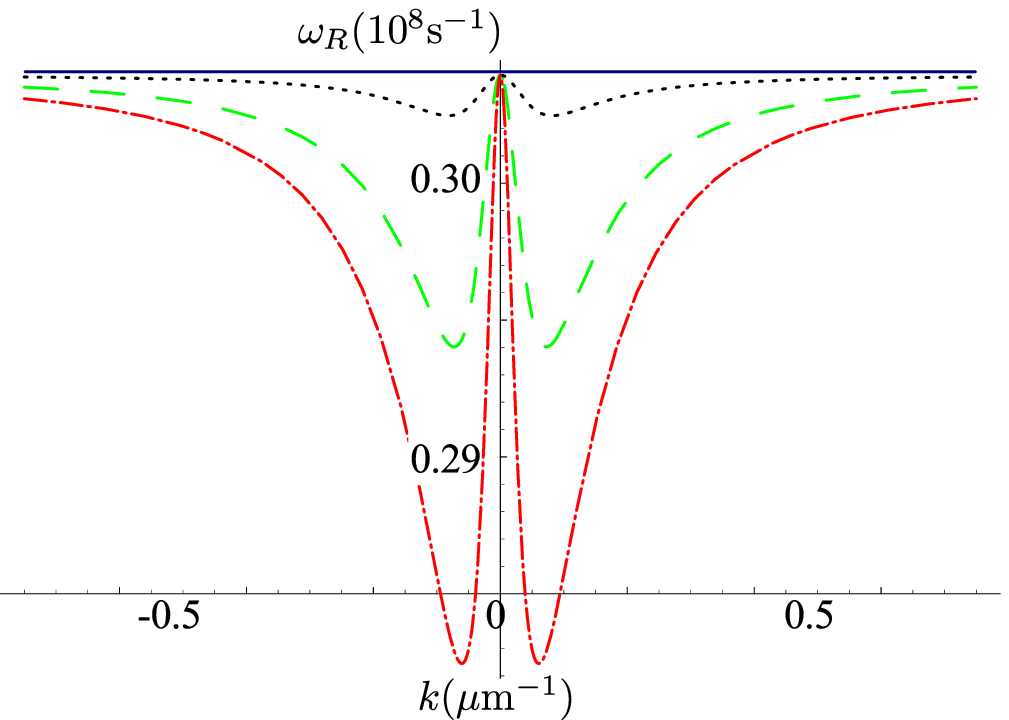}
\caption{(Color online) The dependence of the Rabi frequency $\omega_R$ on  the wave vector $k$ along the CNT (26,20) axis. The amplitude of the ac field is  1 mV/nm. The magnetic field along the CNT axis is equal to $B=0.93 B_z^{cr}$ (dash-dotted), $B= 0.95 B_z^{cr}$ (dashed),  $B= 0.98 B_z^{cr}$ (dotted), and $B=B_z^{cr}$ (solid).}
\label{fig_Rabi_frequency_around_critical_field}
\end{figure}

\begin{figure}[!hbt]
\centering
\includegraphics[width=\linewidth]{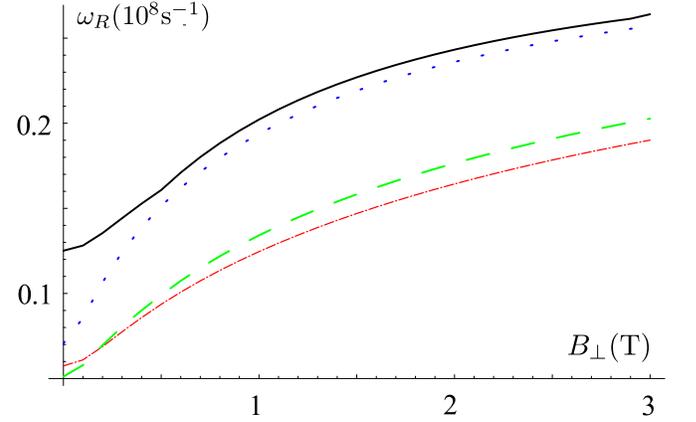}
\caption{(Color online) The dependence of the Rabi frequency $\omega_R$ on the magnetic field  $B_\perp$ applied perpendicular to the CNT axis for metalic nanotubes: the CNT (13,10) -dotted, the CNT (23,20) - dashed, the CNT (33,30) - full. The amplitude of the ac field is  1  mV/nm.}
\label{fig_Rabi_magnetic_field_perpendicular_axis_metal}
\end{figure}

A magnetic field along the nanotube axis renormalizes the coefficients $\beta$ and $\Delta k^t_{\rm cv}$ [see Eqs. (\ref{eq:beta}) and (\ref{eq:k_cv})] and thus allows us to change the Rabi and resonance frequencies. If the magnetic field is chosen such that $k^t+ \Delta k^{t*}=0$, i.e., for $B=B_z^{cr}$ [see Eq. (\ref{compensate_orbital_gap})], with
\begin{equation}
B_z^{cr}= \tau  \frac{\Phi_0}{\pi \hbar v_F}\frac{5.4{\rm meV}}{R[{\rm nm}]^3}\cos3\theta.
\label{compensate_orbital_gap_1},
\end{equation}
the Rabi frequency is $\omega_{R}^{*}$ for arbitrary $k$. Thus, by chosing $B_z$ appropriately, the Rabi frequency can be increased to its upper limit $\omega_R^*$. This effect is stable with respect to small deviations from the optimal $B_z$, as is shown in Fig. \ref{fig_Rabi_frequency_around_critical_field}. Note that for $k=0$, $\omega_R = \omega_R^*$ for any $B_z$ sufficiently close to $B_z^{cr}$.

As usual, a magnetic field $B$ perpendicular to the nanotube axis aligns the electron spin. An ac electric field along this magnetic field induces Rabi transitions. Assuming that the frequency of the electric field is tuned to the energy of the spin splitting at the Fermi points (the Fermi level is assumed to be tuned into the $k=0$ splitting), the dependence of the Rabi frequency on the magnetic field is shown in Fig. \ref{fig_Rabi_magnetic_field_perpendicular_axis_metal}.

If a circularly polarized field is applied, one induces the transitions only from the spin-down state at the Fermi level to the state spin-up above the Fermi level
\begin{equation}
H^{circ}(t) = \tau e E_{circ}(t) \xi S^+ \sigma_2,
\label{E_along_OY_circ}
\end{equation}
with $S^\pm=S^x\pm i S^y$. If the opposite polarization of the field is chosen, the transitions occur in the opposite directions.

\subsection{With static electric field}
A static electric field perpendicular to the axis of a nanotube in combination with SOI lifts the spin degeneracy in the spectrum of an armchair nanotube. In contrast to the well-known Rabi resonance method, realized in a static magnetic field and a perpendicular time-dependent magnetic field, we propose an all-electric setup for spin manipulation with two perpendicular electric fields, one of which is static and the other is time-dependent. The static electric field aligns the spin along the $y$ direction, i.e., perpendicular to the CNT axis and perpendicular to the direction of the static electric field. The time-dependent electric field rotates spin around the $x$ direction.

For armchair nanotubes we find that the transitions between states 1 and 2 or between 3 and 4 in the spectrum shown in Fig.
\ref{fig_helical_modes_10} have the optimal Rabi frequencies
$\omega_R^*$. Rabi transitions between these groups (e.g.
$1\leftrightarrow4$) are not possible.

\begin{figure}[!hbt]
\centering
\includegraphics[width=\linewidth]{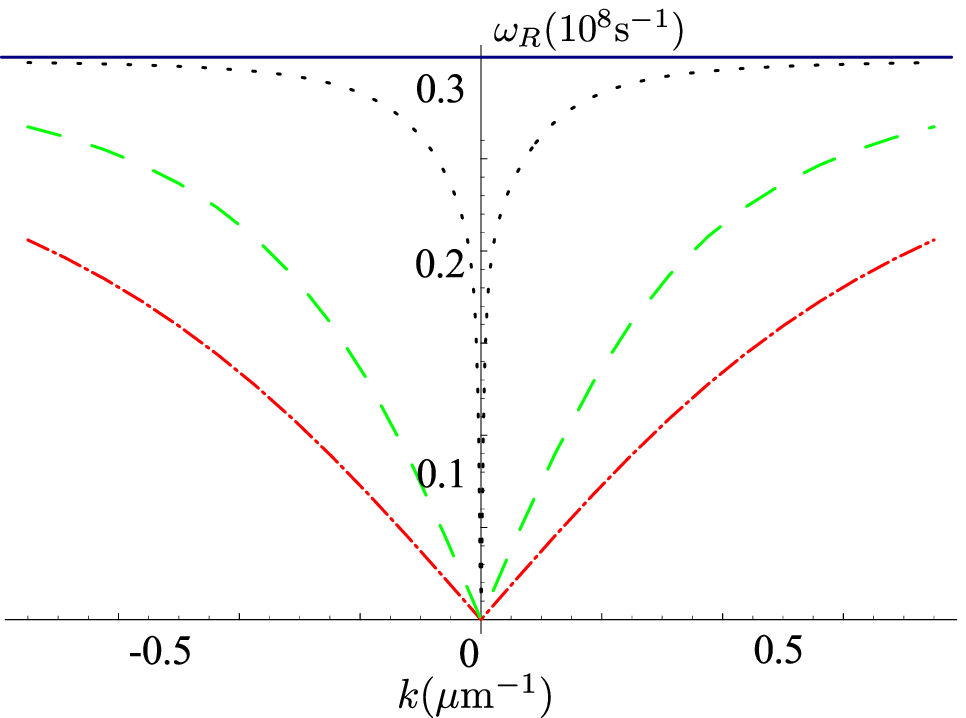}
\caption{(Color online) The dependence of the Rabi frequency $\omega_R$ on  the wave vector $k$ along the CNT (26,20) axis. The amplitude of the ac field is $E_{ac} = 1 mV/nm$ and $E= 1 V/nm$. The magnetic field along the CNT axis is equal to $B=0$ (dash-dotted), $B= 0.5 B_z^{cr}$ (dashed),  $B= 0.9 B_z^{cr}$ (dotted), and $B=B_z^{cr}$ (solid).}
\label{fig_with_electric_field_Rabi_frequency_around_critical_field}
\end{figure}

As explained above, for non-armchair but metallic CNTs, the orbital term $\Delta k^{t*}$ can be compensated by an additional magnetic field $B_z=B_z^{cr}$ along the CNT. Then, as in the case of the armchair nanotube, the Rabi frequency of transitions between states 1 and 2 or between 3 and 4 is $\omega_{R}^{*}$, while transitions between states of different groups are not allowed. For $B_z\neq B_z^{cr}$, the Rabi frequencies are smaller than $\omega_{R}^{*}$ and depend on $k$, as is shown in Fig. \ref{fig_with_electric_field_Rabi_frequency_around_critical_field}.

A magnetic field perpendicular to the nanotube axis (see Fig.\ref{fig_Magnetic_Field_Along_Electric_Field}) breaks the symmetry of the spectrum around the Dirac point. As a result, the resonance frequencies for the right-moving and left-moving modes are different and it is possible to implement the EDSR mechanism for only one of the two modes.

\section{Conclusions\label{section_conclusions}}

We have studied the interplay of strong electric fields, magnetic fields and spin-orbit interactions in carbon nanotubes. An approximate effective low-energy theory describing the electrons near the two Dirac points has been derived analytically and this theory has been tested against more sophisticated numerical solutions of the lattice tight-binding Hamiltonian for the second shell $\pi$ and $\sigma$ orbitals. We have established that the properties of carbon nanotubes are described well by our analytical model in the limit of large field screening. The latter can be achieved by immersing the CNT into dielectrica.

The central feature of CNTs in electric fields is the appearance of (spin-filtered) helical modes in an all-electric setup. For perfect armchair nanotubes, there are two pairs of helical modes, one for each valley, transporting up-spins in one direction and down-spins in the opposite direction. This helicity is perfect in that the average spin is non-zero only for this one spin component and zero for all others. Thus, the average spin is a perfectly odd function of $k$ for armchair CNTs. For non-armchair chiralities, an additional magnetic field can be used to restore the helical phase in one valley. In the other valley, all electronic states are removed from the low-energy regime so that this valley is suppressed by the combination of non-armchair chirality and magnetic field.

Furthermore, we have shown that the EDSR effect may be implemented by a time-dependent electric field perpendicular to the CNT. The typical Rabi frequencies which can be achieved in this system are in the MHz-GHz range.

\acknowledgments

We acknowledge helpful comments by P. Recher and A. Schultes. This work was partially supported by the Swiss NSF, NCCR Nanoscience, NCCR QSIT, and DARPA.

\bibliography{references_manuel}

\end{document}